\documentclass[journal]{IEEEtran}

\usepackage{amsmath,amstext,amsfonts,amssymb,eucal,graphicx}
\usepackage{subfigure}
\usepackage{epsfig}

\newtheorem{theorem}{Theorem}
\newtheorem{definition}{Definition}

\newcommand{\sluttlinje}{\hspace*{1cm}\hspace*{-1cm}~\hfill}
\newcommand{\sluttmerke}{\sluttlinje\raisebox{-1mm}{\rule{2.5mm}{2.5mm}}}
\newcommand{\freestar}{ \framebox[7pt]{$\star$} }

\newcommand{\boxslash}{\begin{picture}(9,5)  \put(2,0){\framebox(5,5){$\smallsetminus$}} \end{picture}}

\font\BBb = dsrom10

\newcommand{\C}{{\BBb C}}
\newcommand{\Z}{{\BBb Z}}

\begin{document}
\bibliographystyle{IEEEtran}
\title{Free deconvolution for signal processing applications}
\author{\O yvind Ryan,~\IEEEmembership{Member,~IEEE,}
                                M{\'e}rouane~Debbah,~\IEEEmembership{Member,~IEEE}
\thanks{\O yvind~Ryan is with the  Mobile Communications Group, Institut Eurecom, 2229 Route des Cretes, B.P. 193, 06904 SOPHIA ANTIPOLIS CEDEX,
        and Department of Informatics, Group for Digital Signal Processing and Image Analysis, University of Oslo, Gaustadalleen 23, P.O Box 1080 Blindern, NO-0316 Oslo, oyvindry@ifi.uio.no}
\thanks{M{\'e}rouane~Debbah is with the  Mobile Communications Group, Institut Eurecom, 2229 Route des Cretes, B.P. 193, 06904 SOPHIA ANTIPOLIS CEDEX, debbah@eurecom.fr}}

\markboth{IEEE Transactions on Information Theory,~Vol.~1,
No.~1,~January~2007}{Shell \MakeLowercase{\textit{et al.}}: Bare
Demo of IEEEtran.cls for Journals}

\maketitle
\begin{abstract}
Situations in many fields of research, such as digital
communications, nuclear physics and mathematical finance, can be
modelled with random matrices. When the matrices get large, free
probability theory is an invaluable tool for describing the
asymptotic behaviour of many systems. It will be shown how free
probability can be used to aid in source detection for certain
systems. Sample covariance matrices for systems with noise are the
starting point in our source detection problem. Multiplicative
free deconvolution is shown to be a method which can aid in
expressing limit eigenvalue distributions for sample covariance
matrices, and to simplify estimators for eigenvalue distributions
of covariance matrices.
\end{abstract}

\begin{keywords}
Free Probability Theory, Random Matrices, deconvolution, limiting
eigenvalue distribution, MIMO, G-analysis.
\end{keywords}

\section{Introduction}
Random matrices, and in particular limit distributions of sample
covariance matrices, have proved to be a useful tool for modelling
systems, for instance in digital communications
\cite{paper:telatar99,paper:tsehanly,paper:shamai01,paper:verdu99},
nuclear physics~\cite{paper:guhr,book:mehta} and mathematical
finance~\cite{book:bouchaud,paper:gallucio}. A typical random
matrix model is the information-plus-noise model,
\begin{equation} \label{system1}
  {\bf W}_n = \frac{1}{N}({\bf R}_n + \sigma {\bf X}_n)({\bf R}_n + \sigma {\bf X}_n)^{H}.
\end{equation}
${\bf R}_n$ and ${\bf X}_n$ are assumed independent random
matrices of dimension $n\times N$ throughout the paper, where
${\bf X}_n$ contains i.i.d. standard (i.e. mean $0$, variance $1$)
complex Gaussian entries. (\ref{system1}) can be thought of as the
sample covariance matrices of random vectors ${\bf r}_n + \sigma
{\bf x}_n$. ${\bf r}_n$ can be interpreted as a vector containing
the system characteristics (direction of arrival for instance in
radar applications or impulse response in channel estimation
applications). ${\bf x}_n$ represents additive noise, with
$\sigma$ a measure of the strength of the noise. Throughout the
paper, $n$ and $N$ will be increased so that
\begin{equation} \label{gcondition}
  \lim_{n\rightarrow\infty}\frac{n}{N} = c,
\end{equation}
i.e. the number of observations is increased at the same rate as
the number of parameters of the system. This is typical of many
situations arising in signal processing applications where one can
gather only a limited number of observations during which the
characteristics of the signal do not change.

The situation motivating our problem is the following: Assume that
$N$ observations are taken by $n$ sensors. Observed values at each
sensor may be the result of an unknown number of sources with
unknown origins. In addition, each sensor is under the influence
of noise. The sensors thus form a random vector ${\bf r}_n+\sigma
{\bf x}_n$, and the observed values form a realization of the
sample covariance matrix ${\bf W}_n$. Based on the fact that ${\bf
W}_n$ is known, one is interested in inferring as much as possible
about the random vector ${\bf r}_n$, and hence on the system
(\ref{system1}). Within this setting, one would like  to connect
the following quantities:
\begin{enumerate}
  \item the eigenvalue distribution of ${\bf W}_n$,
  \item the eigenvalue distribution of ${\bf \Gamma}_n=\frac{1}{N}{\bf R}_n{\bf R}_n^{H}$,
  \item the eigenvalue distribution of the covariance matrix ${\bf \Theta}_n= E\left({\bf r}_n{\bf r}_n^{H} \right)$.
\end{enumerate}
In~\cite{paper:doziersilverstein1}, Dozier and Silverstein explain
how one can use 2) to estimate 1) by solving a given equation.
However, no algorithm for solving it was provided. In fact, many
applications are interested in going from 1) to 2) when attempting
to retrieve information about the system. Unfortunately,
~\cite{paper:doziersilverstein1} does not provide any hint on this
direction. Recently, in~\cite{eurecom:multfreeconv},  we show that
the framework of~\cite{paper:doziersilverstein1} is an
interpretation of the concept of  {\em multiplicative free
convolution}. Moreover,~\cite{eurecom:multfreeconv} introduces the
concept of free deconvolution and  provides an estimate of 2) from
1) in a similar way as estimating 1) from 2).

3) can be adressed by the {\em
$G_2$-estimator}~\cite{chapter:girkotenyears}, which provides a
consistent estimator for the Stieltjes transform of covariance
matrices, the basis for the estimation being the Stieltjes
transform of sample covariance matrices. $G$-estimators have
already shown their usefulness in many
applications~\cite{paper:mestre} but still lack  intuitive
interpretations. In~\cite{eurecom:multfreeconv}, we also show that
the $G_2$-estimator can be derived within the framework of
multiplicative free convolution. This provides a computational
algorithm   for finding 2). Note  that 3) can be found directly,
without finding 2) as demonstrated in~\cite{paper:raoedelman}.
However, the latter does not provide a unified framework for
computing the complete eigenvalue distribution but only a set of
moments.

Beside the mathematical framework, we also address implementation
issues of free deconvolution. Interestingly,  multiplicative free
deconvolution admits a convenient implementation, which will be
described and demonstrated in this paper. Such an implementation
will be used to address several problems related to signal
processing. For communication systems, estimation of the rank of
the signal subspace, noise variance and channel capacity will be
addressed.

This paper is organized as follows. Section~\ref{framework}
presents the basic concepts needed on free probability, including
multiplicative and additive free convolution and deconvolution.
Section~\ref{sectionsystem1} states the results for systems of
type (\ref{system1}). In particular, finding quantities 2) and 3)
from  quantity 1) will be addressed here.
Section~\ref{frameworkimpl}
presents implementation issues  of these concepts.
Section~\ref{apps} will explain through examples and simulations
the importance of the system (\ref{system1})
for digital communications. In the following, upper (lower
boldface) symbols will be used for matrices (column vectors)
whereas lower symbols will represent scalar values, $(.)^T$ will
denote transpose operator, $(.)^\star$ conjugation and
$(.)^H=\left((.)^T\right)^\star$ hermitian transpose. ${\bf I}$
will represent the identity matrix.

\section{Framework for free convolution} \label{framework}
Free probability~\cite{book:hiaipetz} theory has grown into an
entire field of research through the pioneering work of Voiculescu
in the 1980's~\cite{vo2}~\cite{paper:vomult}~\cite{vo6}~\cite{vo7}. The basic
definitions of free probability are quite abstract, as the aim was
 to introduce  an analogy to independence in classical
probability that can be used for non-commutative random variables
like matrices. These more general random variables are elements in
what is  called a  {\em noncommutative probability space}. This
can be defined by a pair $(A,\phi)$, where $A$ is a unital {\em
$\ast$-algebra} with unit $I$, and $\phi$ is a normalized (i.e.
$\phi(I)=1$) linear functional on $A$. The elements of $A$ are
called random variables. In all our examples, $A$ will consist of
$n\times n$ matrices or random matrices. For matrices, $\phi$ will
be the normalized trace $tr_n$, defined by (for any $a \in A$)
\[
  tr_n(a) = \frac{1}{n} Tr(a) = \frac{1}{n} \sum_{i=1}^n a_{ii},
\]
while for random matrices, $\phi$ will be the linear functional $\tau_n$ defined by
\[
  \tau_n(a) = \frac{1}{n} \sum_{i=1}^n E(a_{ii}) = E(tr_n(a)).
\]

 The unit in these $\ast$-algebras is the
$n\times n$ identity matrix ${\bf I_n}$. The noncommutative probability
spaces considered will all be {\em tracial}, i.e. $\phi$
satisfies the trace property $\phi(ab) = \phi(ba)$. The analogy to
independence is called freeness:
\begin{definition} \label{freedef}
  A family of unital $\ast$-subalgebras
  $(A_i)_{i\in I}$ will be called a free family if
\begin{equation} \label{freeeq}
       \left\lbrace  \begin{matrix}
           a_j\in A_{i_j} \\
           i_1\neq i_2,i_2\neq i_3,\cdots ,i_{n-1}\neq i_n \\
           \phi(a_1)=\phi(a_2)=\cdots =\phi(a_n)=0 \end{matrix}
       \right\rbrace
  \Rightarrow \phi(a_1\cdots a_n)=0.
\end{equation}
  A family of random variables $a_i$ is called a free family if the algebras they generate form a free family.
\end{definition}

One can note  that the condition $i_1\neq i_n$ is not included in
the definition of freeness. This may seem strange since if $\phi$
is a trace and $i_1=i_n$, we can rearrange the terms so that two
consecutive terms in (\ref{freeeq}) come from the same algebra. If
this rearranged term does not evaluate to zero through the
definition of freeness, the definition of freeness would be
inconsistent. It is not hard to show that this small issue does
not cause an inconsistency problem. To see this, assume that
(\ref{freeeq}) is satisfied for all indices where the circularity
condition $i_1\neq i_n$ is satisfied. We need to show that
(\ref{freeeq}) also holds for indices where $i_1=i_n$.
By writing
\begin{equation} \label{splitting1}
  a_na_1 = (a_na_1 -\phi(a_na_1)I) + \phi(a_na_1)I = b_1 + \phi(a_na_1)I,
\end{equation}
we can express $\phi(a_1\cdots a_n) = \phi(a_na_1 a_2 \cdots a_{n-1})$
as a sum of the two terms
$\phi(b_1 a_2\cdots a_{n-1})$ and $\phi(a_na_1)\phi(a_2\cdots a_{n-1})$.
The first term is $0$ by assumption, since $\phi(b_1) = 0$, $b_1\in A_{i_n}$ and $i_n\neq i_{n-1}$.
The second term $\phi(a_na_1) \phi(a_2\cdots a_{n-1})$ contributes with zero when $i_2\neq i_{n-1}$ by assumption.
If $i_2=i_{n-1}$, we use the same splitting as in (\ref{splitting1}) again,
but this time on $\phi(a_2\cdots a_{n-1})=\phi(a_{n-1}a_2 a_3\cdots a_{n-2})$,
to conclude that $\phi(a_2\cdots a_{n-1})$ evaluates to zero unless $i_3=i_{n-2}$.
Continuing in this way, we will eventually arrive at
the term $\phi( a_{n/2} a_{n/2+1} )$ if $n$ is even,
or the term $\phi(a_{(n+1)/2})$ if $n$ is odd.
The first of these is $0$ since $i_{n/2} \neq i_{n/2+1}$, and the second is $0$ by assumption.

\begin{definition}
We will say that a sequence of random variables $a_{n1},a_{n2},...$ in probability spaces $(A_n,\phi_n)$ converge in distribution if,
for any $m_1,...,m_r\in\Z$, $k_1,...,k_r\in\{ 1,2,...\}$,
we have that the limit $\phi_n( a_{nk_1}^{m_1}\cdots a_{nk_r}^{m_r})$ exists as $n\rightarrow\infty$.
If these limits can be written as $\phi( a_{k_1}^{m_1}\cdots a_{k_r}^{m_r})$ for some noncommutative probability space $(A,\phi)$ and free random variables $a_1,a_2,... \in (A,\phi)$,
we will say that the $a_{n1},a_{n2},...$ are {\em asymptotically free}.
\end{definition}

Asymptotic freeness is a very useful concept for our purposes,
since many types of random matrices exhibit asymptotic freeness
when their sizes get large. For instance, consider random matrices
$\frac{1}{\sqrt{n}} {\bf A}_{n1},\frac{1}{\sqrt{n}}{\bf
A}_{n2},...$, where the ${\bf A}_{ni}$ are $n\times n$ with all
entries independent and standard Gaussian (i.e. mean $0$ and
variance $1$). Then it is well-known~\cite{book:hiaipetz} that the
$\frac{1}{\sqrt{n}}{\bf A}_{ni}$ are asymptotically free. The
limit distribution of the $\frac{1}{\sqrt{n}} {\bf A}_{ni}$ in
this case is called {\em circular}, due to the asymptotic
distribution of the eigenvalues of $\frac{1}{\sqrt{n}}{\bf
A}_{ni}$: When $n\rightarrow\infty$, these get uniformly
distributed inside the unit circle of the complex plane
\cite{paper:girko84,paper:bai97}.

(\ref{freeeq}) enables us to calculate the mixed moments of free
random variables $a_1$ and $a_2$. In particular, the moments of
$a_1+a_2$ and $a_1a_2$ can be calculated. In order to calculate
$\phi((a_1+a_2)^4)$, multiply out $(a_1+a_2)^4$, and use linearity
and (\ref{freeeq}) to calculate all
$\phi(a_{i_1}a_{i_2}a_{i_3}a_{i_4})$ ($i_j = 1$, $2$). For
example, to calculate $\phi(a_1a_2a_1a_2)$, write it as
\[
  \begin{split}
    \phi( &\left( (a_1 - \phi(a_1)I) + \phi(a_1)I \right) \left( (a_2 - \phi(a_2)I) + \phi(a_2)I \right) \\
               &\left( (a_1 - \phi(a_1)I) + \phi(a_1)I \right) \left( (a_2 - \phi(a_2)I) + \phi(a_2)I \right) ),
  \end{split}
\]
and multiply it out as $16$ terms. The term
\[
  \begin{split}
    \phi( &(a_1 - \phi(a_1)I) (a_2 - \phi(a_2)I) \\
               &(a_1 - \phi(a_1)I) (a_2 - \phi(a_2)I) )
  \end{split}
\]
is zero by (\ref{freeeq}).
The term
\[
  \begin{array}{ll}
     & \phi( (a_1 - \phi(a_1)I) \phi(a_2)I (a_1 - \phi(a_1)I) (a_2 - \phi(a_2)I) )\\
     & = \phi(a_2) \phi( (a_1 - \phi(a_1)I) (a_1 - \phi(a_1)I) (a_2 - \phi(a_2)I) )
  \end{array}
\]
can be calculated by writing
\[
  b = (a_1 - \phi(a_1)I) (a_1 - \phi(a_1)I)
\]
(which also is in the algebra generated by $a_1$),
setting
\[
  b = (b-\phi(b)I) + \phi(b)I,
\]
and using (\ref{freeeq}) again. The same procedure can be followed for any mixed moments.

When the sequences of moments uniquely identify probability measures (which is the case for compactly supported probability measures),
the distributions of $a_1+a_2$ and $a_1a_2$ give us two new probability measures, which depend only on the
probability measures associated with the moments of $a_1$, $a_2$.
Therefore we can define two operations on the set of probability measures:
{\em Additive free convolution}
\begin{equation} \label{addconvdef}
  \mu_1\boxplus\mu_2
\end{equation}
for the sum of free random variables, and {\em multiplicative free convolution}
\begin{equation} \label{multconvdef}
  \mu_1\boxtimes\mu_2
\end{equation}
for the product of free random variables. These operations can be
used to predict the spectrum of sums or products of asymptotically
free random matrices. For instance, if $a_{1n}$ has an eigenvalue
distribution which approaches $\mu_1$ and $a_{2n}$ has an
eigenvalue distribution which approaches $\mu_2$, one has that the
eigenvalue distribution of $a_{1n}+a_{2n}$ approaches
$\mu_1\boxplus\mu_2$, so that $\mu_1\boxplus\mu_2$ can be used as
an eigenvalue predictor for large matrices. Eigenvalue prediction
for combinations of matrices is in general not possible, unless we
have some assumption on the eigenvector structures. Such an 
assumption which makes random matrices fit into a free probability
setting (and make therefore the random matrices free), is that of
{\em uniformly distributed eigenvector structure} (i.e. the
eigenvectors point in some sense in all directions with equal
probability).

We will also find it useful to introduce the concepts of {\em
additive and multiplicative free deconvolution}:
\begin{definition}
  Given probability measures $\mu$ and $\mu_2$.
  When there is a unique probability measure $\mu_1$ such that
  \[
    \mu = \mu_1 \boxplus \mu_2 \mbox{, } \mu = \mu_1 \boxtimes \mu_2 \mbox{ respectively,}
  \]
  we will write
  \[
    \mu_1 = \mu \boxminus \mu_2 \mbox{, } \mu_1 = \mu \boxslash \mu_2 \mbox{ respectively.}
  \]
  We say that $\mu_1$ is the additive free deconvolution (respectively multiplicative free deconvolution) of $\mu$ with $\mu_2$.
\end{definition}
It is noted that the symbols presented here for additive and
multiplicative free deconvolution have not been introduced in the
literature previously. With additive free deconvolution, one can
show that there always is a unique $\mu_1$ such that $\mu = \mu_1
\boxplus \mu_2$. For multiplicative free deconvolution, a unique
$\mu_1$ exists as long as we assume non-vanishing first moments of
the measures. This will always be the case for the measures we
consider.

\begin{figure}
  \begin{center}
    \epsfig{figure=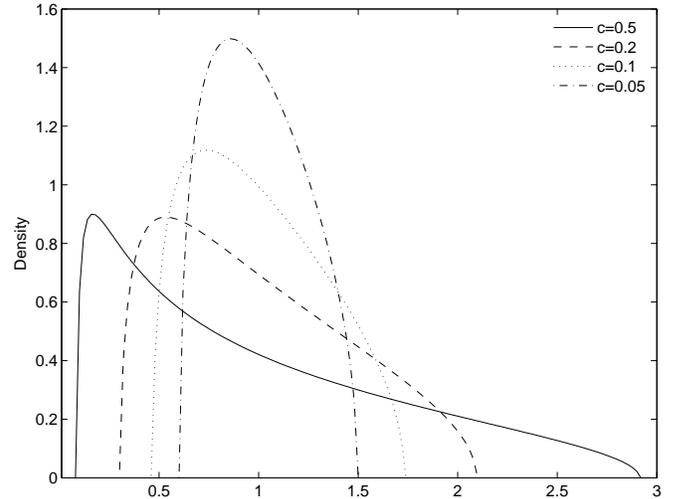,width=0.99\columnwidth}
  \end{center}
  \caption{Different Mar\u{c}henko Pastur laws $\mu_c$.}\label{fig:simmpplot}
\end{figure}

Some probability measures appear as limits for large random
matrices in many situations. One important measure is the
Mar\u{c}henko Pastur law $\mu_c$ (\cite{book:tulinoverdu} page 9),
also known as the free Poisson distribution in free probability.
It is characterized by the density
\begin{equation} \label{mpdensity}
  f^{\mu_c}(x) = (1-\frac{1}{c})^+ \delta(x) + \frac{\sqrt{(x-a)^+(b-x)^+}}{2\pi cx},
\end{equation}
where $(z)^+ =\mbox{max}(0,z)$, $a=(1-\sqrt{c})^2$ and
$b=(1+\sqrt{c})^2$.
In figure~\ref{fig:simmpplot}, $\mu_c$ is plotted for some values of $c$.
It is known that $\mu_c$ describes asymptotic eigenvalue
distributions of {\em Wishart} matrices. Wishart matrices have the
form $\frac{1}{N} {\bf R}{\bf R}^{H}$, where ${\bf R}$ is an
$n\times N$ random matrix with independent standard Gaussian
entries. $\mu_c$ appears as limits of such when
$\frac{n}{N}\rightarrow c$ when $n\rightarrow\infty$,  Note that
the Mar\u{c}henko Pastur law can also hold in the limit for
non-gaussian entries.

We would like to describe free convolution in terms of the
probability densities of the involved measures, since this
provides us with the eigenvalue distributions. An important tool
in this setting is {\em the Stieltjes transform}
(\cite{book:tulinoverdu} page 38). For a probability measure
$\mu$, this is the analytic function on $\C^+ = \{ z\in C : Im z >
0 \}$ defined by
\begin{equation} \label{st1}
  m_{\mu}(z) = \int \frac{1}{\lambda-z} dF^{\mu}(\lambda),
\end{equation}
where $F^{\mu}$ is the cumulative distribution function of $\mu$.
All $\mu$ we will consider have support on the positive part of
the real line. For such $\mu$, $m_{\mu}$ can be analytically
continued to the negative real line, where the values of $m_{\mu}$
are real. If $\mu$ has compact support, we can expand $m_{\mu}(z)$
in a Laurent series, where the coefficient are the moments $\mu_k$
of $\mu$:
\begin{equation} \label{st2}
  m_{\mu}(z) = -\frac{1}{z} \sum_{k=0}^{\infty} \frac{\mu_k}{z^k}.
\end{equation}
A convenient inversion formula for the Stieltjes transform also
exists: We have
\begin{equation} \label{st3}
  f^{\mu}(\lambda) = \lim_{\omega\rightarrow 0^+} \frac{1}{\pi} Im \left[ m_{\mu}(\lambda + j\omega) \right].
\end{equation}

\section{Information plus noise model} \label{sectionsystem1}
In this section we will indicate how the quantities  2) and 3) can
be found. The connection between the information plus noise model
and free convolution will be made.

\subsection{Estimation of the sample covariance matrix 2)}
In~\cite{eurecom:multfreeconv}, the following result was shown.
\begin{theorem} \label{teo1d}
  Assume that ${\bf \Gamma}_n=\frac{1}{N}{\bf R}_n{\bf R}_n^{H}$ converge in distribution almost surely to a compactly supported probability measure $\mu_{\Gamma}$.
  Then we have that ${\bf W}_n$ also converge in distribution almost surely to a compactly supported probability measure $\mu_W$ uniquely identified by
  \begin{equation} \label{starteq}
    \mu_W \boxslash \mu_c = (\mu_{\Gamma} \boxslash \mu_c) \boxplus \mu_{\sigma^2 I}.
  \end{equation}
\end{theorem}

Theorem~\ref{teo1d} addresses the relationship from 2) to 1),
since (\ref{starteq}) can be "deconvolved" to the following form:

\begin{equation}
  \mu_W = \left( (\mu_{\Gamma} \boxslash \mu_c) \boxplus \mu_{\sigma^2 I} \right) \boxtimes \mu_c.
\end{equation}

It also addresses the relationship from 1) to 2), which is of
interest here, through deconvolution in  the following form:
\begin{equation} \label{step1}
  \mu_{\Gamma} = \left( (\mu_W \boxslash \mu_c) \boxminus \mu_{\sigma^2 I} \right) \boxtimes \mu_c.
\end{equation}

\subsection{Estimation of the covariance matrix 3)}
General statistical analysis of observations, also called {\em
$G$-analysis}~\cite{book:girkostat}~\cite{paper:mestre} is a
mathematical theory studying complex systems, where the number of
parameters of the considered mathematical model can increase
together with the growth of the number of observations of the
system. The mathematical models which in some sense approach the
system are called {\em $G$-estimators}, and the main difficulty in
$G$-analysis is to find consistent $G$-estimators. We use $N$ for
the number of observations of the system, and $n$ for the number
of parameters of the mathematical model. The condition used in
$G$-analysis expressing the growth of the number of observations
vs. the number of parameters in the mathematical model, is called
the {\em $G$-condition}. The $G$-condition used throughout this
paper is (\ref{gcondition}).

We restrict our analysis to systems where a number of independent
random vector observations are taken, and where the random vectors
have identical distributions. If a random vector has length $n$,
we will use the notation ${\bf \Theta}_n$ to denote the
covariance. Girko calls an estimator for the Stieltjes transform
of covariance matrices a  $G_2$-estimator. In chapter 2.1
of~\cite{chapter:girkotenyears} he introduces the following
expression as candidate for a $G_2$-estimator:
\begin{equation} \label{gestcondition0}
  G_{2,n}(z) = \frac{\hat{\theta}(z)}{z} m_{\mu_{\Gamma_n}}(\hat{\theta}(z)),
\end{equation}
where the term $m_{\mu_{\Gamma_n}}(\hat{\theta}(z))=n^{-1}
Tr\left\{ {\bf \Gamma}_n -\hat{\theta}(z){\bf I}_n \right\}^{-1}$.
The function $\hat{\theta}(z)$ is the solution to the equation.
\begin{equation} \label{gestcondition}
  \hat{\theta}(z) c m_{\mu_{\Gamma_n}}(\hat{\theta}(z)) - (1-c) + \frac{\hat{\theta}(z)}{z} = 0.
\end{equation}
Girko claims that a function $G_{2,n}(z)$ satisfying
(\ref{gestcondition}) and (\ref{gestcondition0}) is a good
approximation for the Stieltjes transform of the involved
covariance matrices $m_{\mu_{\Theta_n}}(z) = \frac{1}{n} Tr\left\{
{\bf \Theta}_n - z {\bf I}_n \right\}^{-1}$. He shows that, for
certain values of $z$, $G_{2,n}(z)$ gets close to
$m_{\mu_{\Theta_n}}(z)$ when $n\rightarrow\infty$.

In~\cite{eurecom:multfreeconv}, the following connection between
the $G_2$-estimator and multiplicative free convolution is made:
\begin{theorem} \label{teo2}
  For the $G_2$-estimator given by (\ref{gestcondition0}), (\ref{gestcondition}), the following holds for a nonempty open set in $\C^+$:
  \begin{equation}
    G_{2,n}(z) = m_{\mu_{\Gamma_n} \boxslash \mu_c}
  \end{equation}
\end{theorem}

Theorem~\ref{teo2} shows that multiplicative free convolution can
be used to estimate the covariance of systems. This addresses the
problem of estimating   quantity 3). Hence, the estimation of
quantity 2) and 3) can be combined, since (\ref{step1}) can be
rewritten to
\begin{equation} \label{step12}
  \mu_{\Gamma} \boxslash \mu_c = (\mu_W \boxslash \mu_c) \boxminus \mu_{\sigma^2 I}.
\end{equation}

Therefore, the following procedure needs to take place to estimate
quantity 3)
\begin{itemize}
  \item[1]- perform multiplicative free deconvolution  of  the measure associated with ${\bf W}_n$  and  the marchenko pastur law  using  the
  $G_2$-estimator.
  \item[2]- perform additive free deconvolution with $\mu_{\sigma^2 I}$. In other words, 
    perform a shift of the spectrum.
\end{itemize}

In section~\ref{sim2}, these steps are performed in the setting of
channel correlation estimation. The combinatorial implementation
of free deconvolution from section~\ref{strategy1} is used to
compute the $G_2$-estimator.

\begin{figure}
  \subfigure[$N=64$]{\epsfig{figure=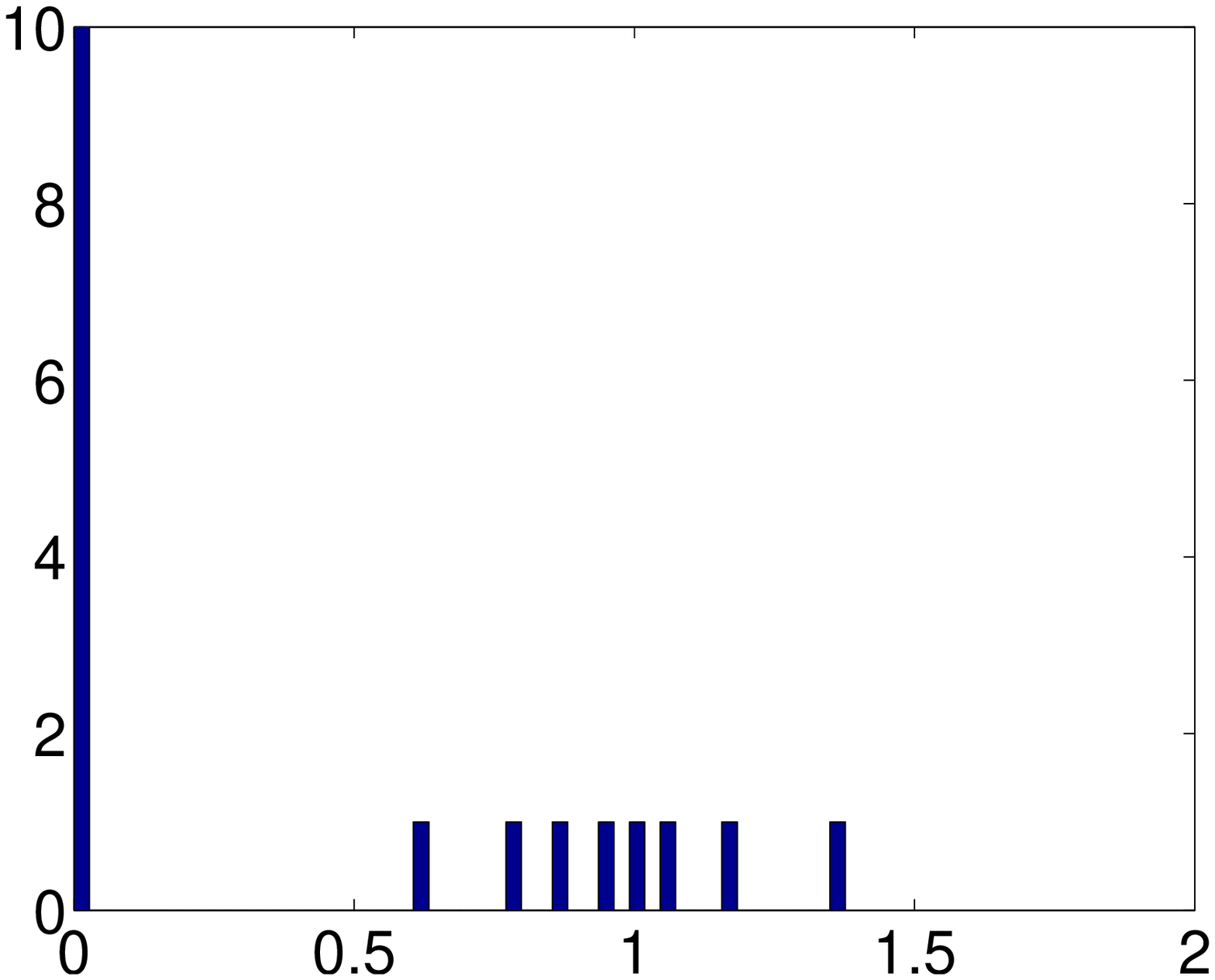,width=0.49\columnwidth}}
  \subfigure[$N=512$]{\epsfig{figure=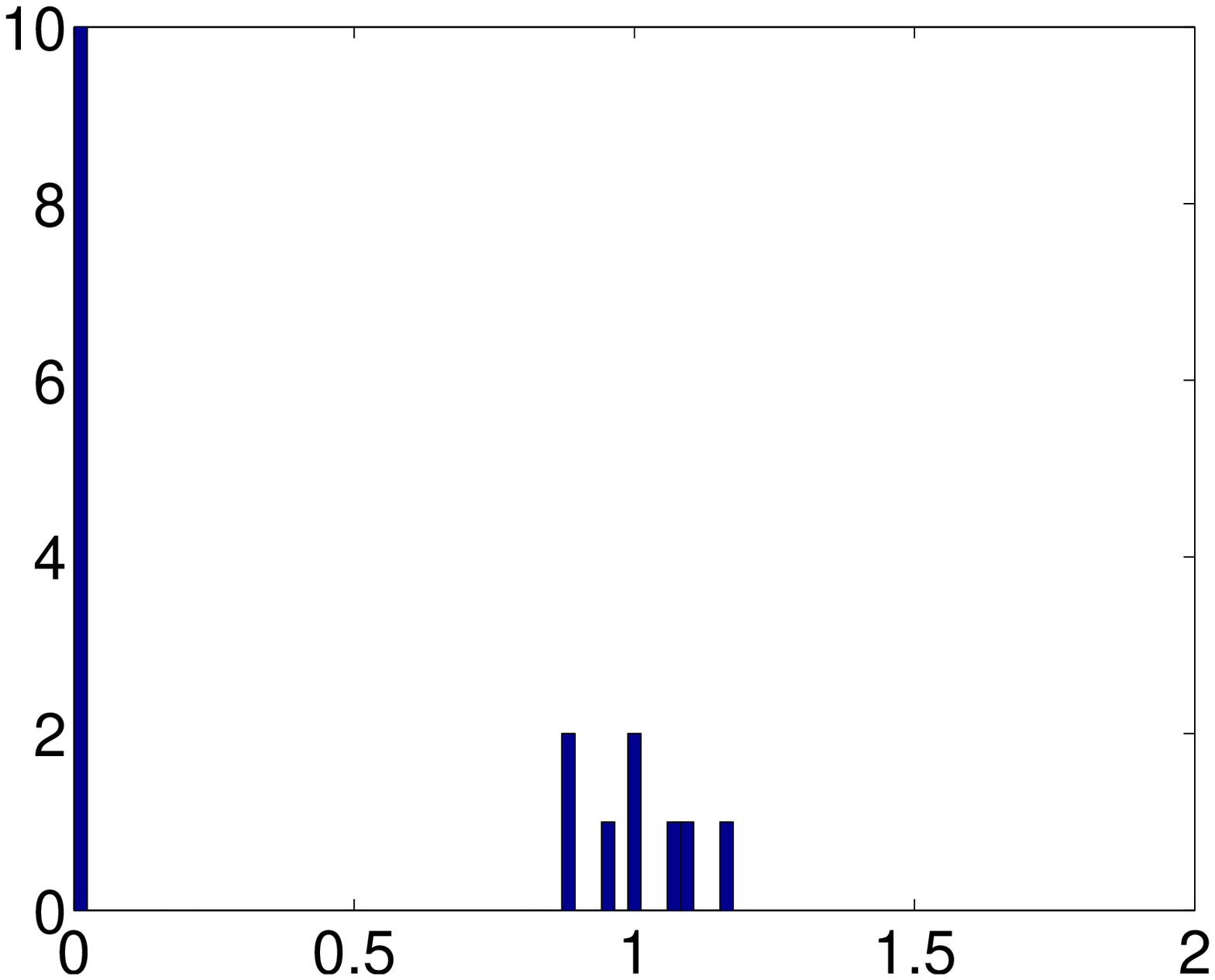,width=0.49\columnwidth}}
  \caption{Histograms of eigenvalues of sample covariance matrices.
    The covariance matrix has rank $K=8$. We choose different number of sensors $N$, and choose $c=0.5$
    (so that $L=2N$ observations are taken).}\label{fig:simsensors}
\end{figure}

We plot in figure~\ref{fig:simsensors} histograms of eigenvalues
for various sample covariance matrices when the rank is $K=8$. As
one can see, if the number of sensors ($N$) are chosen much larger
than the number of signals $K$, the eigenvalues corresponding to
the signals only make up a small portion of the entire set of
eigenvalues. If one has information on the number of impinging
signals, it can therefore be wise to choose the appropriate number
of sensors.

In this paper, the difference between a probability measure, $\mu$,
and an estimate of it, $\nu$,
will be measured in terms of the {\em Mean Square Error of the moments} (MSE).
If the moments of $\int x^k d\mu(x)$, $\int x^k d\nu(x)$ are denoted by $\mu_k$, $\nu_k$, respectively,
the MSE is defined by
\begin{equation} \label{msedef}
  \sum_{k\leq n} | \mu_k - \nu_k |^2
\end{equation}
for some number $n$.
In our estimation problems, the measure $\nu$ we test which gives the minimum MSE (MMSE) will be chosen.

The measure $\mu$ can either be given explicitly in terms of the
distribution of matrices $R_n$ (for which the measure is discrete
and the moments are given by $\nu_k = tr_n(R_n^k)$), or random
matrices, or it can be given in terms of just the moments. In a
free probability setting, giving just the moments is often
convenient, since free convolution can be viewed as operating on
the moments. Since the $G_2$-estimator uses free deconvolution, it
will be subject to a Mean Square Error of moments analysis. We
will compute the MSE for different number of moments, depending on
the availability of the moments.

\begin{figure}
  \subfigure[4 moments]{\epsfig{figure=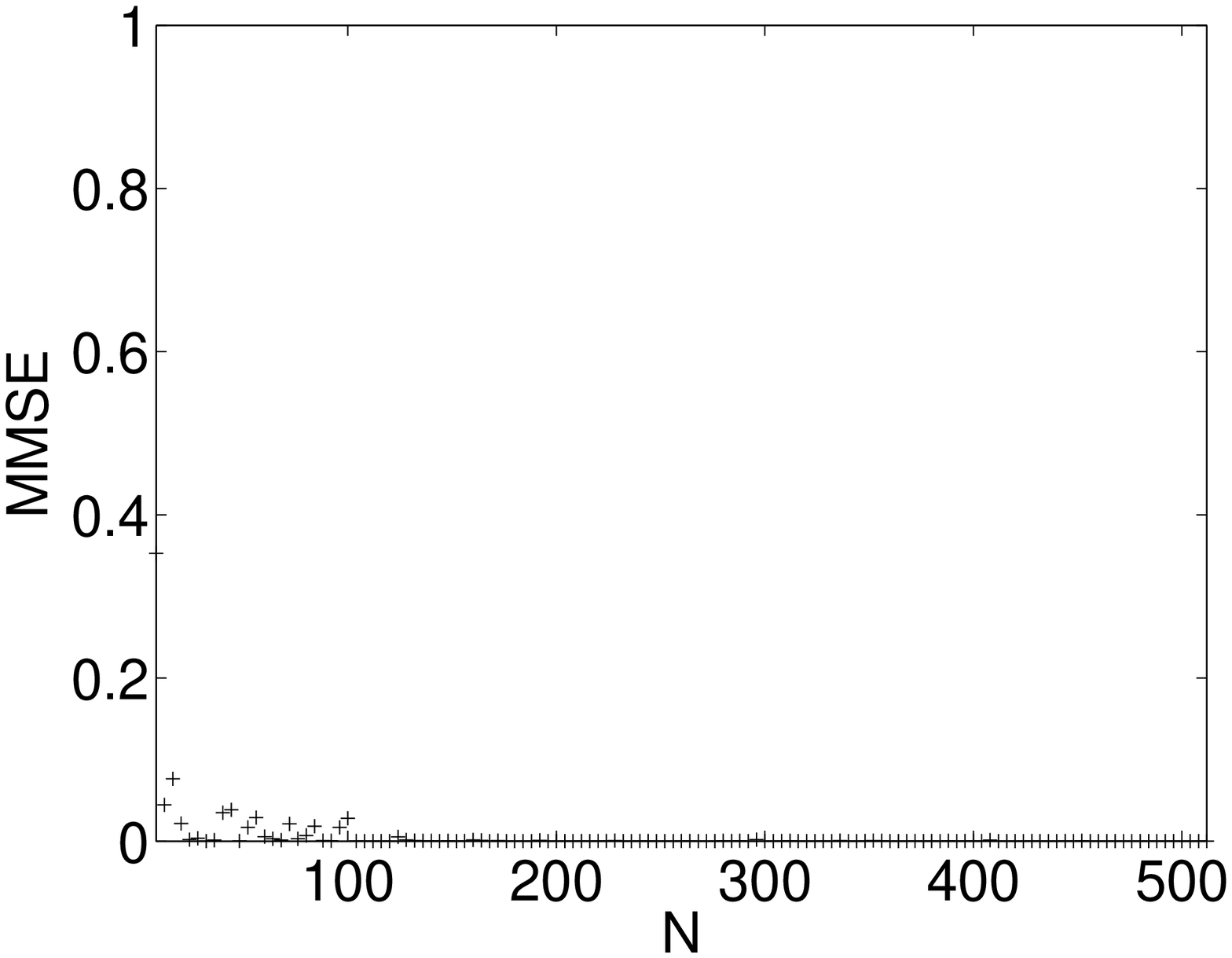,width=0.49\columnwidth}}
  \subfigure[8 moments]{\epsfig{figure=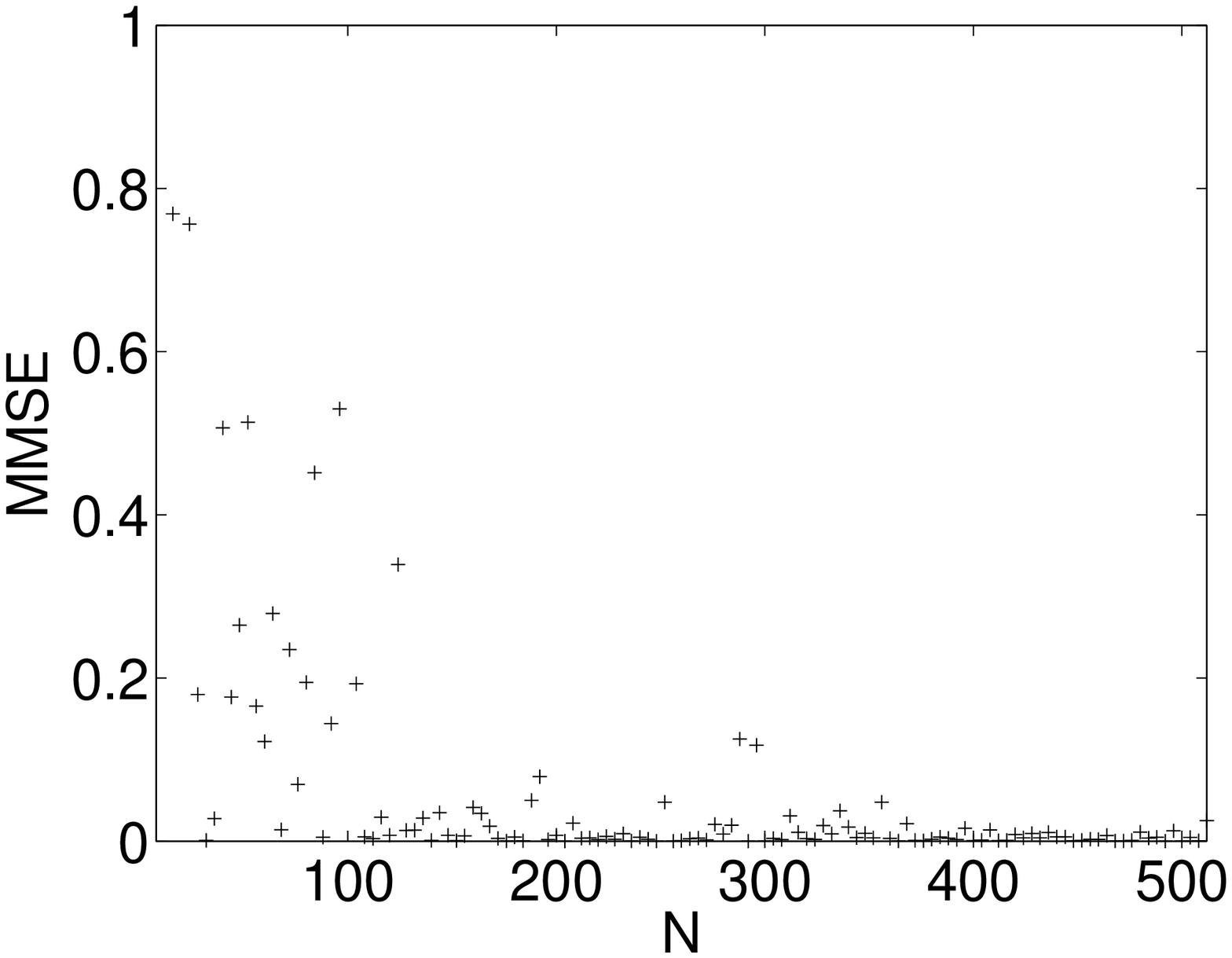,width=0.49\columnwidth}}
  \caption{MMSE of the first moments of the covariance matrices,
    and the first moments of the $G_2$ estimator of the sample covariance matrices.
    The covariance matrices all have distribution $\frac{1}{2}\delta_0 + \frac{1}{2}\delta_1$.
    Different matrix sizes $N$ are tried.
    The value $c=0.5$ is used.}\label{fig:simgirko}
\end{figure}

In figure~\ref{fig:simgirko}, a covariance matrix with density
$\frac{1}{2}\delta_0 + \frac{1}{2}\delta_1$ has been estimated
with the the $G_2$-estimator. Sample covariance matrices of
various sizes are formed, and method A in section~\ref{strategy1}
was used for the free deconvolution in the $G_2$-estimator.
Finally, MSE of the first $4$ and $8$ moments were computed. It is
seen that the MSE decreases with the matrix sizes, which confirms
the accuracy of the $G_2$-estimator. Also, the MSE is much higher
when more moments are included. This is as expected, when we
compare known exact expressions for moments of 
Gaussian random matrices~\cite{paper:thorbjornsen1}, with the 
limits these converge to. 

Since the MSE is typically higher in the higher moments, 
we will in the simulations in this paper minimize a {\em weighted MSE}:
\begin{equation} \label{wmsedef}
   \sum_k w_k | \mu_k - \nu_k |^2,
\end{equation}
instead of the MSE in (\ref{msedef}). Higher moments should have
smaller weights $w_k$, since errors in random matrix models are
typically larger in the higher moments. There may not be optimal
weights which work for all cases. For the cases in this paper, the 
weights  are $w_{2k} = \left( \stackrel{2k}{k} \right)$ and
$w_{2k+1} = 0$, whick coincide with the Catalan numbers $C_k$. 
These weights are motivated from formulas for moments of Gaussian random matrices 
in the way explained below, and are used in this paper since the models we consider 
often involve Gaussian matrices. 

Moment $2k$ of a standard selfadjoint
Gaussian matrix $X_n$ of size $n\times n$
satisfies~\cite{book:hiaipetz}~\cite{book:comblect}
\[
  \lim_{n\rightarrow\infty} \tau_n( {\bf X}_n^{2k} ) = C_k.
\]
Also, exact formulas for $\tau_n( {\bf X}_n^{2k} )$ (4.1.19
in~\cite{book:hiaipetz}) exist, where the proof uses combinatorics
and noncrossing partitions (see section~\ref{strategy1}), with
emphasis on the quantity $C_k$ being
the number of noncrossing partitions of $\{ 1,...,2k\}$ where all
blocks have cardinality two. From the exact formula for the moment
of $\tau_n( {\bf X}_n^{2k} )$, one can also see the difference
between the limit as $n\rightarrow\infty$. This difference is
approximately $n^{-1} \times card(S)$, where $S$ is another set of
partitions. Although this set of partitions is not the same as the
noncrossing partitions, it can in some sense be thought of as
"partitions where limited crossings may be allowed". The choice of
$C_k$ as weight is merely motivated from the belief that $S$ 
possibly has similar properties as the
noncrossing partitions, and that possibly the cardinality has a
similar expression.

In summary, figure~\ref{fig:simgirko} shows that for large matrices,
the $G_2$-estimator gets close to the actual covariance
matrices although several sources can give contribution to errors:
\begin{enumerate}
  \item The sample covariance matrix itself is estimated,
  \item the estimator itself contributes to the error,
  \item the implementation of free deconvolution also contributes to the error.
\end{enumerate}

\section{Computation of free convolution} \label{frameworkimpl}
One of the challenges in free probability theory is the practical
computation of free convolution. Usual results exhibit asymptotic
convergence of product and sum of measures, but do not explicitly
provide a framework for computing the result. In this section,
given two compactly supported probability measures, we will sketch
how their free (de)convolved counterparts  can be computed. In the
cases we are interested (signal impaired with noise), the
Mar\u{c}henko Pastur law $\mu_c$ will be one of the operand
measures, while the other will be a discrete measure, i.e. with
density
\begin{equation} \label{ourforminit}
  f^{\mu}(x) = \sum_{i=1}^n p_i\delta_{\lambda_i}(x),
\end{equation}
where  $p_i$ is the mass at $\lambda_i$, and
$\sum_i p_i = 1$. All $\lambda_i \geq 0$, since
only measures with support on the positive real line are
considered ($n\times n$ sample covariance matrices have such
eigenvalue distributions).  This would be the distribution we
observe in a practical scenario: Since a finite number of samples
are taken, we only observe a discrete estimate of the sample
covariance matrix.

 The Stieltjes transform $m_{\mu_c
\boxslash \mu}$ can be found exactly for $z$ on the negative real
line by solving function equations~\cite{eurecom:multfreeconv},
but one has to perform analytical continuation to the upper half
of the complex plane prior to using the Stieltjes inversion
formula. Indeed, note that since the power series (\ref{st2}) is
only known to converge for $z$ outside a disk with radius equal to
the spectral radius, partial sums of (\ref{st2}) can not
necessarily be used to approach the limit in (\ref{st3}). However,
one can show that values of $m_{\mu}(z)$ for $z$ on the negative
real line can be approximated by analytically continuing partial
sums of (\ref{st2}).

When $\mu$ is discrete, one can show that solving the function
equations boils down to finding the roots of real polynomials (see
section~\ref{strategy3}), which then must be analytically
continued. We will sketch a particular case where this can be
performed exactly. We will also sketch two other methods for
computing free convolution numerically. One method uses a
combinatorial description, which easily admits an efficient
recursive implementation. The other method is based on results on
asymptotic freeness of large random matrices.

\subsection{Numerical Methods}
\subsubsection{Method A: Combinatorial computation of free convolution}
\label{strategy1}
The concept we need for computation of free
convolution presented in this section is that of {\em noncrossing
partitions}~\cite{book:comblect}:
\begin{definition}
  A partition $\pi$ is called noncrossing if whenever we have $i<j<k<l$ with $i\sim k$, $j\sim l$ ($\sim$ meaning belonging to the same block),
  we also have $i\sim j\sim k\sim l$ (i.e. $i,j,k,l$ are all in the same block).
  The set of noncrossing partitions of $\{ 1,,,.,n\}$ is denoted $NC(n)$.
\end{definition}
We will write $\pi = \{ B_1,...,B_r \}$ for the blocks of a partition.
$|B_i|$ will mean the cardinality of the block $B_i$.

{\bf Additive free convolution. }

A convenient way of implementing
additive free convolution comes through the {\em moment-cumulant
formula}, which expresses a relationship between the moments of
the measure and the associated $R$-transform
(\cite{book:tulinoverdu} page 48). The $R$-transform has domain of
definition $\C^+$ and can be defined in terms of the Stieltjes
transform as
\begin{equation} \label{rdef1}
  {\cal R}_{\mu}(z) = m_{\mu}^{-1}(-z) - \frac{1}{z}.
\end{equation}
The importance of the $R$-transform comes from the  additivity
property in additive free convolution,
\begin{equation} \label{radditivity}
  {\cal R}_{\mu_1\boxplus\mu_2}(z) = {\cal R}_{\mu_1}(z) + {\cal R}_{\mu_2}(z).
\end{equation}
Slightly different versions of the $R$-transform are encountered
in the literature. The definition (\ref{rdef1}) is
from~\cite{book:tulinoverdu}. In connection with free
combinatorics, another definition is used, namely $R_{\mu}(z) =
z{\cal R}_{\mu}(z)$. Write $R_{\mu}(z) = \sum_n \alpha_n z^n$. The
coefficients $\alpha_n$ are called cumulants. The moment-cumulant
formula says that
\begin{equation} \label{momentcumulant}
 \mu_n = \sum_{\pi = \{ B_1,\cdots ,B_k \}\in NC(n)} \prod_{i=1}^k \alpha_{|B_i|}.
\end{equation}
From (\ref{momentcumulant}) it follows that the first $n$ cumulants can be computed from the first $n$ moments, and vice versa.
Noncrossing partitions have a structure which makes them easy to iterate over in an implementation.
One can show that (\ref{momentcumulant}) can be rewritten to the following form suitable for implementation:
\begin{equation} \label{computableform}
  \mu_n = \sum_{k\leq n} \alpha_k coef_{n-k} \left( (1+\mu_1z+\mu_2z^2+\cdots)^k \right).
\end{equation}
Here $coef_k$ means the coefficient of $z^k$. Computing the
coefficients of $(1+\mu_1z+\mu_2z^2+\cdots)^k$ can be implemented
in terms of a $k$-fold discrete (classical) convolution, so that
the connection between free and classical convolution can be seen
both in terms of concept and implementation.
(\ref{computableform}) can be implemented in such a way
that the $\mu_n$ are calculated from $\alpha_n$, or, the other way
around, the $\alpha_n$ are calculated from the $\mu_n$.
When computing higher moments, (\ref{computableform}) is quite time-consuming, since many
(classical) convolutions of long sequences have to be performed. A
recursive implementation of (\ref{computableform}) was made for
this paper~\cite{eurecom:freeimpl}, and is briefly described in appendix~\ref{sec:appendix0}.
The paper~\cite{fdsigpro} goes through implementation
issues of free convolution in more detail.

Additive free convolution in terms of moments of the involved
measures can therefore beimplemented through the following steps:
Evaluate cumulants using (\ref{computableform}) for the two
measures, add these, and finally evaluate moments using
(\ref{computableform}) also.

{\bf Multiplicative free convolution. }

The combinatorial transform we need for multiplicative free
convolution is that of {\em boxed
convolution}~\cite{book:comblect} (denoted by $\freestar$), which
can be thought of as a convolution operation on formal power
series. The definition uses noncrossing partitions and will not be
stated here. One power series will be of particular importance to
us. {\em The $Zeta$-series} is intimately connected to $\mu_1$ in
that it appears as it's $R$-transform. It is defined by
\[
  Zeta(z) = \sum_i z^i.
\]
$Zeta(z)$ has an inverse under boxed convolution,
\[
  Moeb(z) = \sum_{n=1}^{\infty} (-1)^{n-1} C_{n-1} z^n,
\]
also called the M\"{o}bius series. Here $(C_n)_{n=1}^{\infty}$ is
the sequence of {\em Catalan numbers} (which are known to be
related to the even moments of Wigner matrices). Define the moment
series of a measure $\mu$ by
\[
  M(\mu)(z) = \sum_{k=1}^{\infty} \mu_k z^k = -\frac{1}{z}m_{\mu}(\frac{1}{z}) - 1.
\]
One can show that (\ref{momentcumulant}) is equivalent to $M(\mu) = R(\mu) \freestar Zeta$.
One can in fact show that boxed convolution on power series is the combinatorial perspective of multiplicative free convolution on measures.
Also,
\begin{enumerate}
  \item boxed convolution with the power series $c^{n-1}Zeta$ represents convolution with the measure $\mu_c$,
  \item boxed convolution with the power series $c^{n-1}Moeb$ represents deconvolution with the measure $\mu_c$.
\end{enumerate}
This is formalized as
\[
  M_{\mu\boxtimes\mu_c} = M_{\mu} \freestar (c^{n-1}Zeta),
\]
and can also be rewritten to
\begin{equation} \label{wayconv}
  cM_{\mu\boxtimes\mu_c} = Zeta \freestar (cM_{\mu}),
\end{equation}
It can be shown that this is nothing but the moment-cumulant formula, 
with cumulants replaced by the coefficients of $cM_{\mu}$, 
and moments replaced by the coefficients $cM_{\mu\boxtimes\mu_c}$. 
Therefore, the same computational procedure can be used for passing between moments and cumulants,
as for passing between the moments series of $\mu\boxtimes\mu_c$ and that of $\mu$,
the only difference being the additional scaling of the moments by $c$:
\begin{enumerate}
  \item multiply all input moments by $c$ prior to execution of (\ref{computableform}),
  \item divide all output moments by $c$ after execution of (\ref{computableform}).
\end{enumerate}
The situation for other compactly supported measures than $\mu_c$ follows the same lines, but with $c^{n-1} Zeta$ and $c^{n-1} Moeb$ replaced with other power series.
Convolution and deconvolution with other measures than $\mu_c$ may be harder to implement, due to the particularly simple structure of the $Zeta$ series.

In addition to computing (\ref{computableform}) and performing step 1) and 2) above, we need first to obtain the moments of $\mu$ in some way.
For $\mu$ as in (\ref{ourforminit}) the moments can be calculated by incrementally computing the numbers $(\lambda_1^m,...,\lambda_n^m)$,
adding these together and normalizing.
At the end, we may also need to retrieve the probability density from the computed moments.
If a density corresponds to the eigenvalue distribution of some matrix,
the {\em Newton-Girard Formulas}~\cite{book:programmingmath} can be used to retrieve the eigenvalues from the moments.
These formulas state a relationship between the elementary symmetric polynomials
\begin{equation}
  \Pi_j(\lambda_1,...,\lambda_n) = \sum_{i_1 < \cdots < i_j\leq n} \lambda_{i_1}\cdots\lambda_{i_j},
\end{equation}
and the sums of the powers of their variables
\begin{equation}
  S_p(\lambda_1,...,\lambda_n) = \sum_{1\leq i\leq n} \lambda_i^p,
\end{equation}
through the recurrence relation
\begin{equation} \label{newtongirard}
\begin{array}{ll}
  & (-1)^m m \Pi_m(\lambda_1,...,\lambda_n) \\
 +& \sum_{k=1}^m (-1)^{k+m} S_k(\lambda_1,...,\lambda_n) \Pi_{m-k}(\lambda_1,...,\lambda_n)= 0.
\end{array}
\end{equation}
If $S_p(\lambda_1,...,\lambda_n)$ are known for $1\leq p\leq n$, (\ref{newtongirard}) can be used repeatedly to compute
$\Pi_m(\lambda_1,...,\lambda_n)$, $1\leq m\leq n$.

Coefficient $n-k$ in the characteristic polynomial
\[
  (\lambda - \lambda_1)\cdots(\lambda - \lambda_n)
\]
is $(-1)^k \Pi_k(\lambda_1,...,\lambda_n)$, and these can be computed from $S_k(\lambda_1,...,\lambda_n)$ using (\ref{newtongirard}).
Since $S_k(\lambda_1,...,\lambda_n)=nm_k$ (with $m_k$ being the kth moment),
the entire characteristic polynomial can be computed from the moments.
Hence, the eigenvalues can be found also.

In general, the density can not be written as the eigenvalue
distribution of a matrix, but the sketched procedure can still
provide us with an estimate based on the moments. Intuitively, the
approximation should work better when more moments are involved.
The simulations in this paper use the sketched procedure only for
a low number of moments, since mostly discrete measures with few
atoms are estimated. We have thus also avoided issues for solving
higher degree characteristic equations with high precision.

\subsubsection{Method B: Computation of free convolution based on asymptotic
freeness results} \label{strategy2} As mentioned, the
Mar\u{c}henko Pastur law can be approximated by random matrices of
the form ${\bf \Gamma}_n=\frac{1}{N} {\bf R}_n {\bf R}_n^{H}$,
where ${\bf R}_n$ is $n\times N$ with i.i.d. standard Gaussian
entries. It is also known that the product of such a ${\bf
\Gamma}_n$ with a (deterministic) matrix with eigenvalue
distribution $\mu$ has an eigenvalue distribution which
approximates that of $\mu_c \boxtimes \mu$. This is formulated in
free probability as a result on {\em asymptotic freeness} of
certain random matrices with deterministic
matrices~\cite{book:hiaipetz}. Therefore, one can approximate
multiplicative free convolution by taking a sample from a random
matrix ${\bf \Gamma}_n$, multiply it with a deterministic diagonal
matrix with eigenvalue distribution $\mu$, and calculating the
eigenvalue distribution of this product. The deterministic matrix
need not be diagonal. Additive free convolution can be estimated
in the same way by adding ${\bf \Gamma}_n$ and the deterministic
matrix instead of multiplying them.

\begin{figure}
  \subfigure[4 moments]{\epsfig{figure=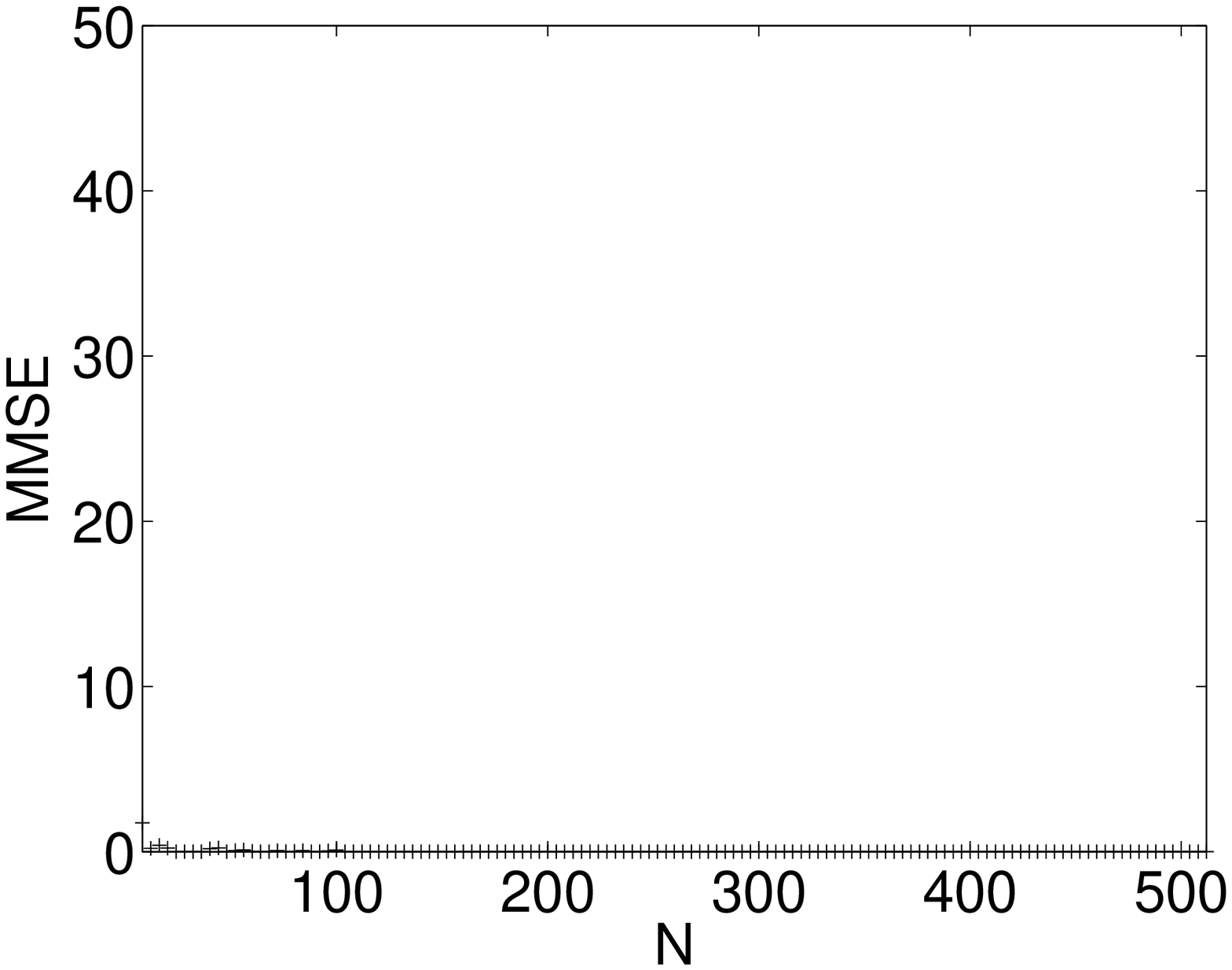,width=0.49\columnwidth}}
  \subfigure[8 moments]{\epsfig{figure=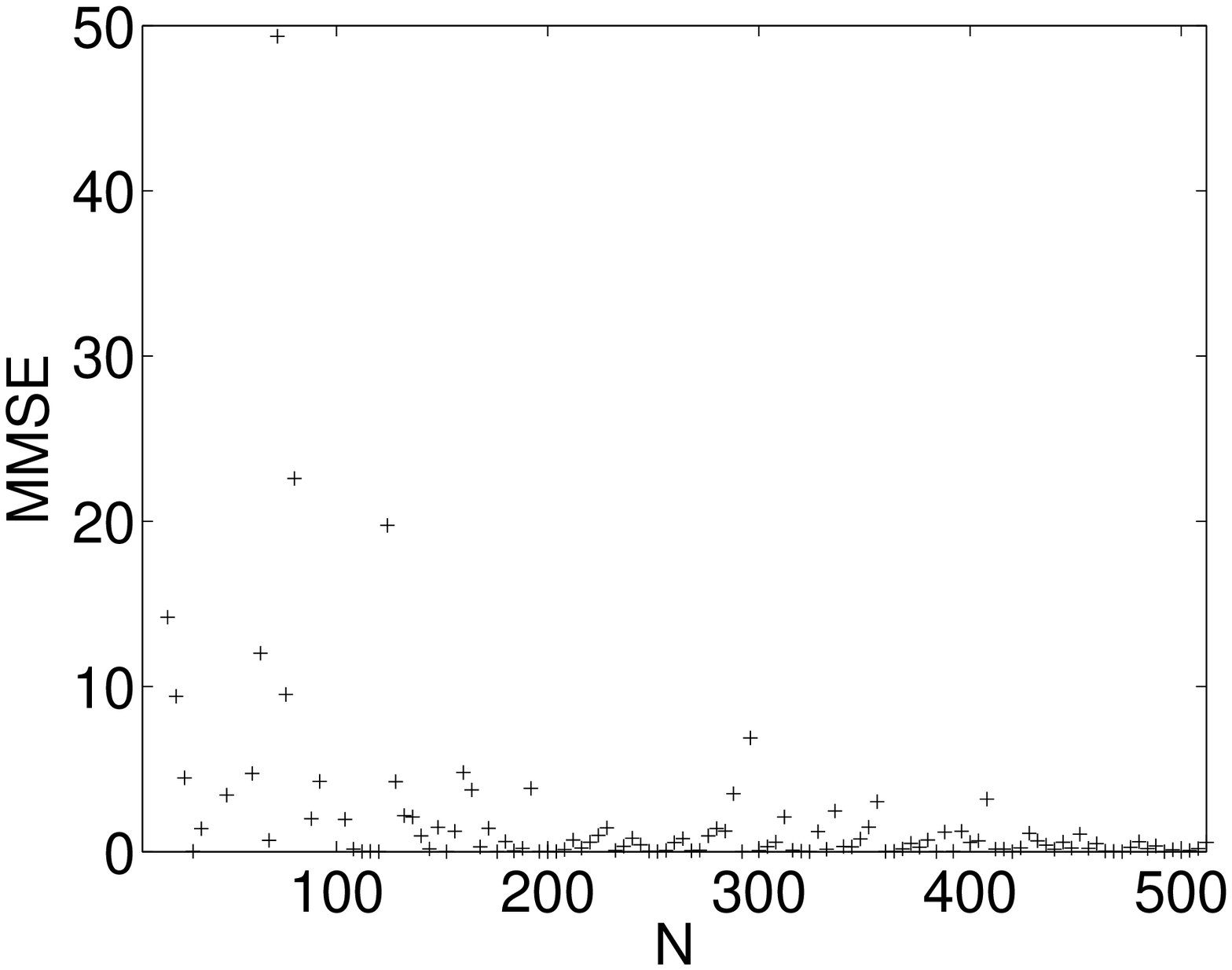,width=0.49\columnwidth}}
  \caption{MMSE of the first moments of $\left( \frac{1}{2}\delta_0 + \frac{1}{2}\delta_1 \right) \boxtimes \mu_c$,
    and the same moments computed approximately with method B using different matrix sizes $N$.
    The value $c=0.5$ is used.}\label{fig:simmethodb}
\end{figure}

In figure~\ref{fig:simmethodb}, method B is demonstrated for various matrix sizes to obtain approximations of
$\left( \frac{1}{2}\delta_0 + \frac{1}{2}\delta_1 \right) \boxtimes \mu_c$ for $c=0.5$.
The moments of the approximations
are compared with the exact moments, which are obtained with method A.
The Mean Square Error of the moments is used to measure the difference.
As in figure~\ref{fig:simgirko}, it is seen that the MSE decreases with the matrix sizes,
and that the MSE is much higher when more moments are included.

\begin{figure}
  \subfigure[$c=0.2$. $L=7680$ observations]{\epsfig{figure=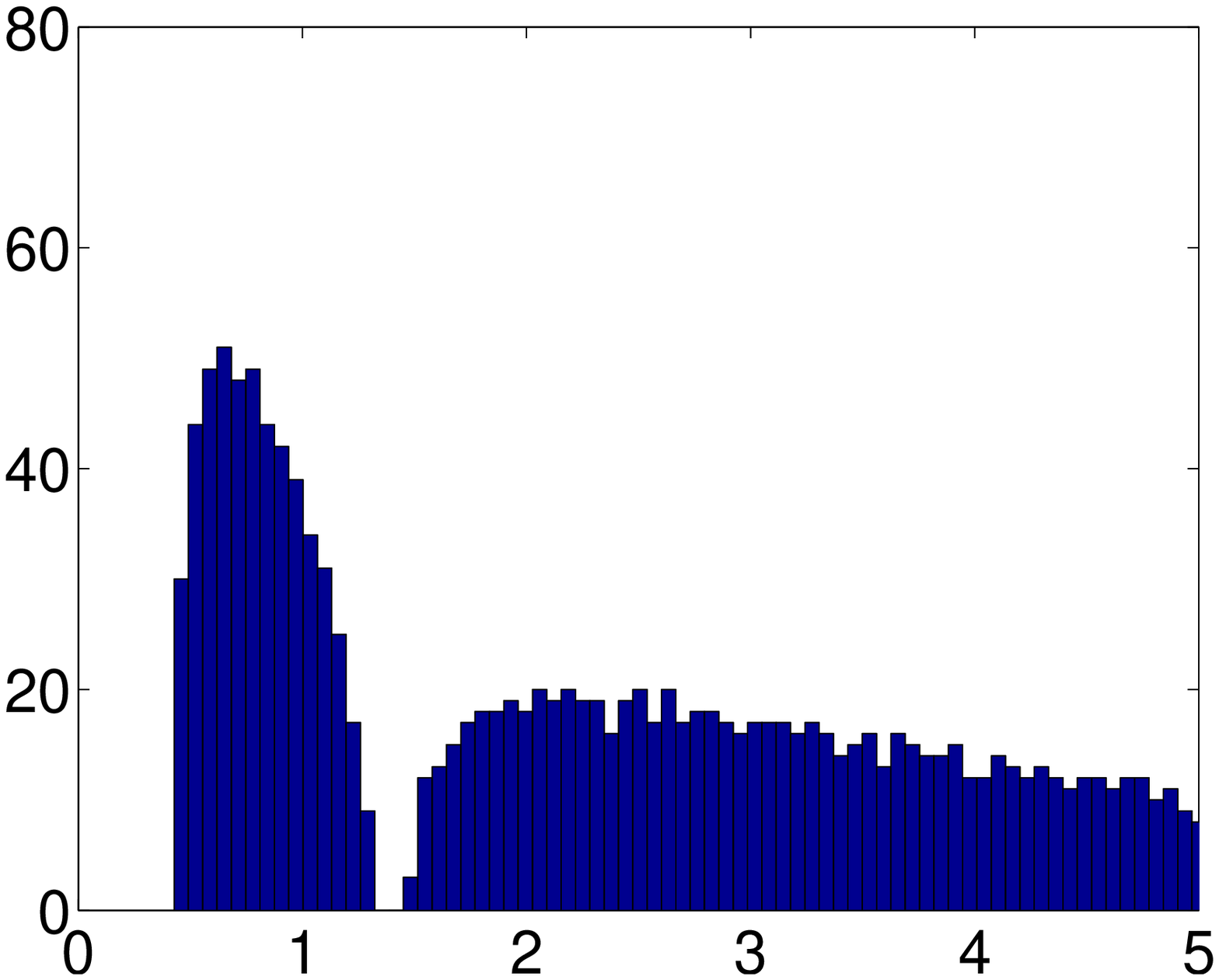,width=0.49\columnwidth}}
  \subfigure[$c=0.05$. $L=30720$ observations]{\epsfig{figure=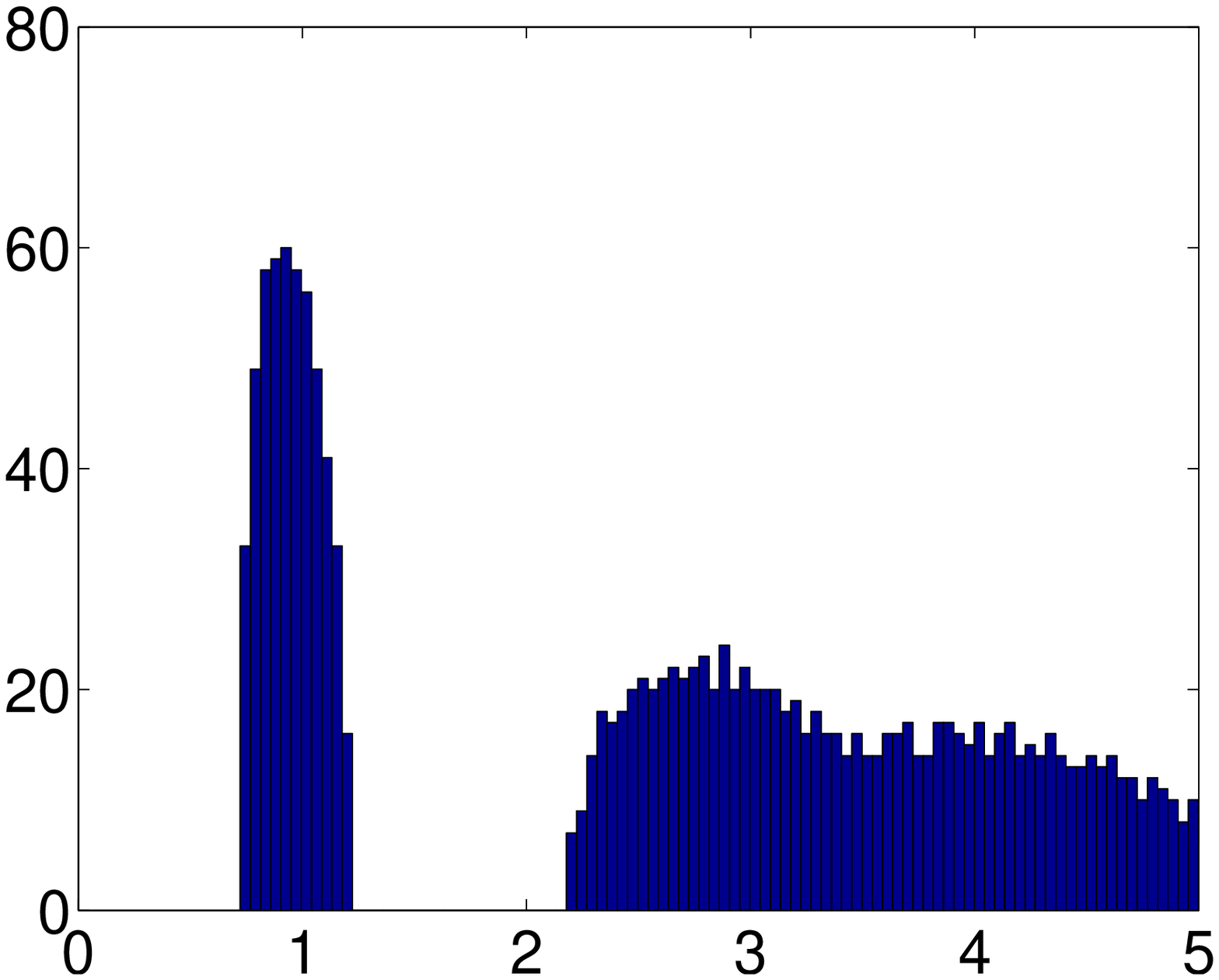,width=0.49\columnwidth}}
  \caption{Approximated densities of $\left( \frac{1}{3}\delta_1(x) + \frac{1}{3}\delta_3(x) + \frac{1}{3}\delta_4(x) \right) \boxtimes\mu_c$
    with method B for various values of $c$}\label{fig:simsplitting}
\end{figure}

Another interesting phenomenon occurs when we let $c$ go to $0$, demonstrated in figure~\ref{fig:simsplitting} for the measure $\mu$ with
\begin{equation} \label{expdist}
  f^{\mu}(x) = \frac{1}{3}\delta_1(x) + \frac{1}{3}\delta_3(x) + \frac{1}{3}\delta_4(x).
\end{equation}
It is seen that for small $c$, the
support of $\mu\boxtimes\mu_c$ seems to split into disjoint components centered at the dirac
locations. This is compatible with results
from~\cite{paper:silversteincombette}. There it is just noted
that, for a given type of systems, the support splits into TWO
different components, and the relative mass between these
components is used to estimate the numbers of signals present in
the systems they consider.
In figure~\ref{fig:simsplitting}, a matrix of dimension $N\times N$ with $N=1536$ and eigenvalue distribution $\mu$ is taken.
This matrix is multiplied with a Wishart matrix $\frac{1}{L} {\bf X} {\bf X}^H$,
where ${\bf X}$ has dimension $N\times L$ with $\frac{N}{L} = c$ with decreasing values of $c$.
It is seen that the dirac at $1$ is split from the rest of the support in both plots, with the split more visible for
the lower value of $c$. The splitting of the two other diracs from each other it not very visible for these values of $c$.
Also, the peaks in the density of $\mu\boxtimes\mu_c$ occur slightly to the left of the dirac points,
which is as expected from the comments succeeding theorem~\ref{exactconv}.

A partial explanation for the fact that $\mu\boxtimes\mu_c$ in some sense converges to $\mu$ when $c\rightarrow 0$
is given by combining the following facts:
\begin{itemize}
  \item The sample covariance matrix converges to the true covariance matrix
    when the number of observations tend to $\infty$ (i.e. $c\rightarrow 0$).
  \item The $G_2$-estimator for the covariance matrices is given by multiplicative free deconvolution with $\mu_c$.
\end{itemize}

In summary, the differences between method A and B are the following:
\begin{enumerate}
  \item Method B needs to compute the full eigenvalue distribution of the operand matrices.
    Method A works entirely on the moments.
  \item With method B, the results are only approximate.
    If the eigenvalues are needed also, method A needs to perform computationally expensive tasks in approximating eigenvalues from moments,
    for instance as described in section~\ref{strategy1}.
  \item Method B is computationally more expensive, in that computations with large matrices are needed in order to obtain accurate results.
    Method A is scalable in the sense that performance scales with the number of moments computed.
    The lower moments are the same regardless on how many higher moments are computed.
\end{enumerate}
The two methods should really be used together: While method A easily can get exact moments,
method B can tell us the accuracy of random matrix approximations by comparison with these exact moments.

The simulations in this paper will use method A, since deconvolution is a primary component, and
since we in many cases can get the results with an MSE of moments analysis. 
Deconvolution with method B should correspond to multiplication with the inverse of a Wishart matrix,
but initial tests do not suggest that this strategy works very well when predicting the deconvolved eigenvalues.

The way method A and method B have been described here, they have in common that they only work for
free convolution and deconvolution with Mar\u{c}henko Pastur laws.
Method A worked since (\ref{computableform}) held for such laws,
while method B worked since these laws have a convenient asymptotic random matrix model in terms of Gaussian random matrices.

\subsection{Non-numerical methods: Exact calculations of multiplicative free convolution}
\label{strategy3}
Computation of free convolution in general has to be performed numerically, for instance through the methods in section~\ref{strategy1} and~\ref{strategy2}.
In some cases, the computation can be performed exactly, i.e. the density of the (de)convolutions can be found exactly.
Consider the specific case of (\ref{ourforminit}) where
\begin{equation} \label{ourform}
  f^{\mu}(x) = (1-p)\delta_0(x) + p\delta_{\lambda}(x),
\end{equation}
where $p<1$, $\lambda > 0$.
Such measures were considered in~\cite{paper:raoedelman},
where deconvolution was implemented by finding a pair $(p,\lambda)$ minimizing the difference between the moments of $\mu\boxtimes\mu_c$,
and the moments of observed sample covariance matrices.
Exact expressions for the density of $\mu\boxtimes\mu_c$ were not used, all calculations were performed in terms of moments.

(\ref{ourform}) contains one dirac (i.e. $p=1$) as a special case.
It is clear that multiplicative free convolution with $\delta_{\lambda}$ has an exact expression,
since we simply multiply the spectrum with $\lambda$ (the spectrum is scaled).
As it turns out, all $\mu$ of the form (\ref{ourform}) give an exact expression for $\mu\boxtimes\mu_c$.
In appendix~\ref{sec:appendixa}, the following is shown:
\begin{theorem} \label{exactconv}
  The density of $\mu\boxtimes\mu_c$ is $0$ outside the interval
  \begin{equation} \label{icp}
    I_{\lambda , c , p} = \left[\lambda(1+cp) - 2\lambda\sqrt{cp}, \lambda(1+cp) + 2\lambda\sqrt{cp} \right],
  \end{equation}
  while the density on $I_{\lambda , c , p}$ is given by
  \begin{equation} \label{density}
    f^{\mu\boxtimes\mu_c}(x) = \frac{ \sqrt{ K_1(x) K_2(x) } }{2c\lambda x\pi},
  \end{equation}
  where
  \[
  \begin{array}{lll}
    K_1(x) &=& x - \lambda(1+cp) + 2\lambda\sqrt{cp}\\
    K_2(x) &=& \lambda(1+cp) + 2\lambda\sqrt{cp} - x.
  \end{array}
  \]
  The density has a unique maximum at $x=\lambda\frac{(1-cp)^2}{1+cp}$, with value $\frac{\sqrt{cp}}{c\pi\lambda(1-cp)}$.
\end{theorem}

\begin{figure}
  \subfigure[$c=0.5$]{\epsfig{figure=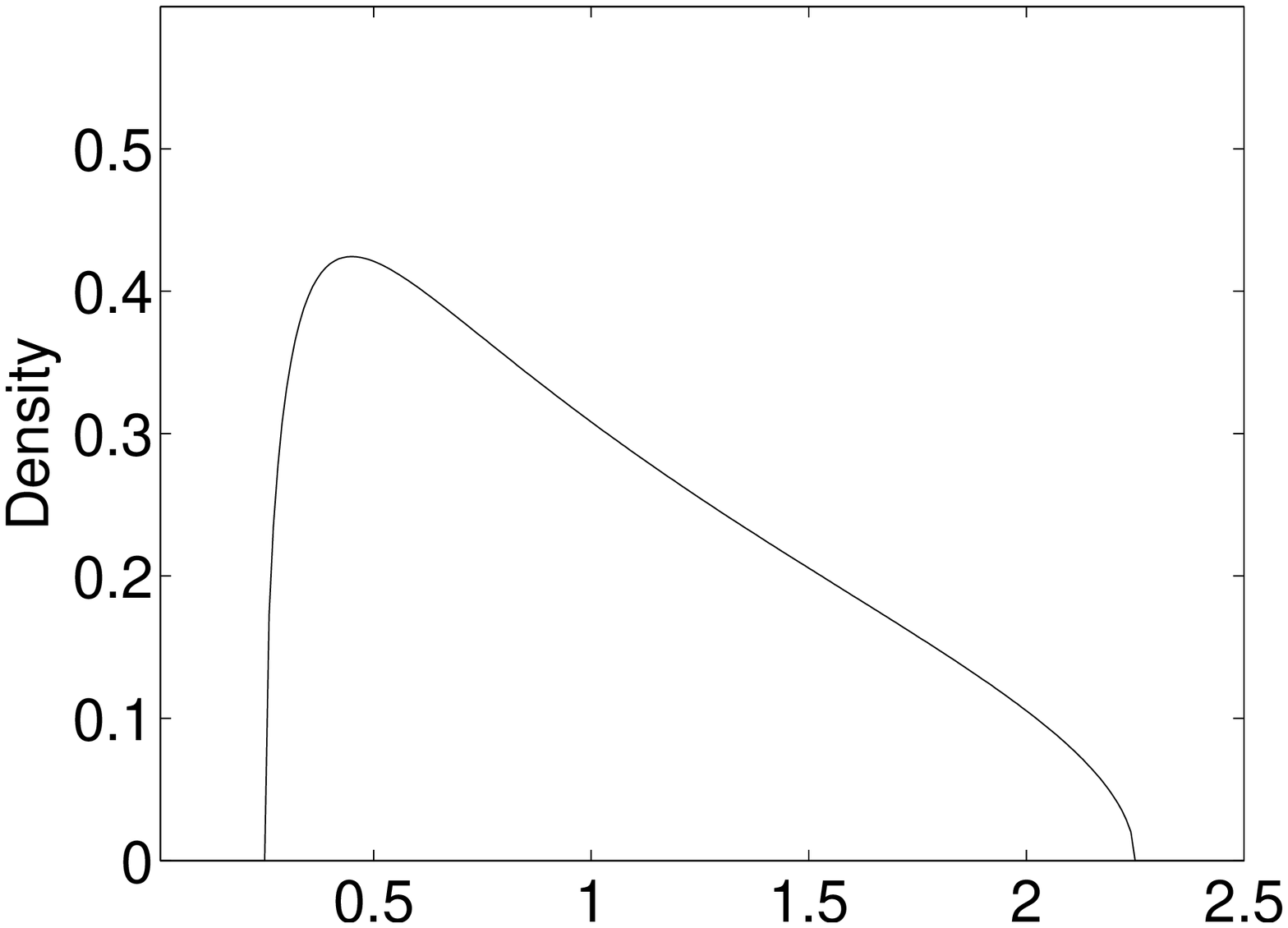,width=0.49\columnwidth}}
  \subfigure[$c=0.25$]{\epsfig{figure=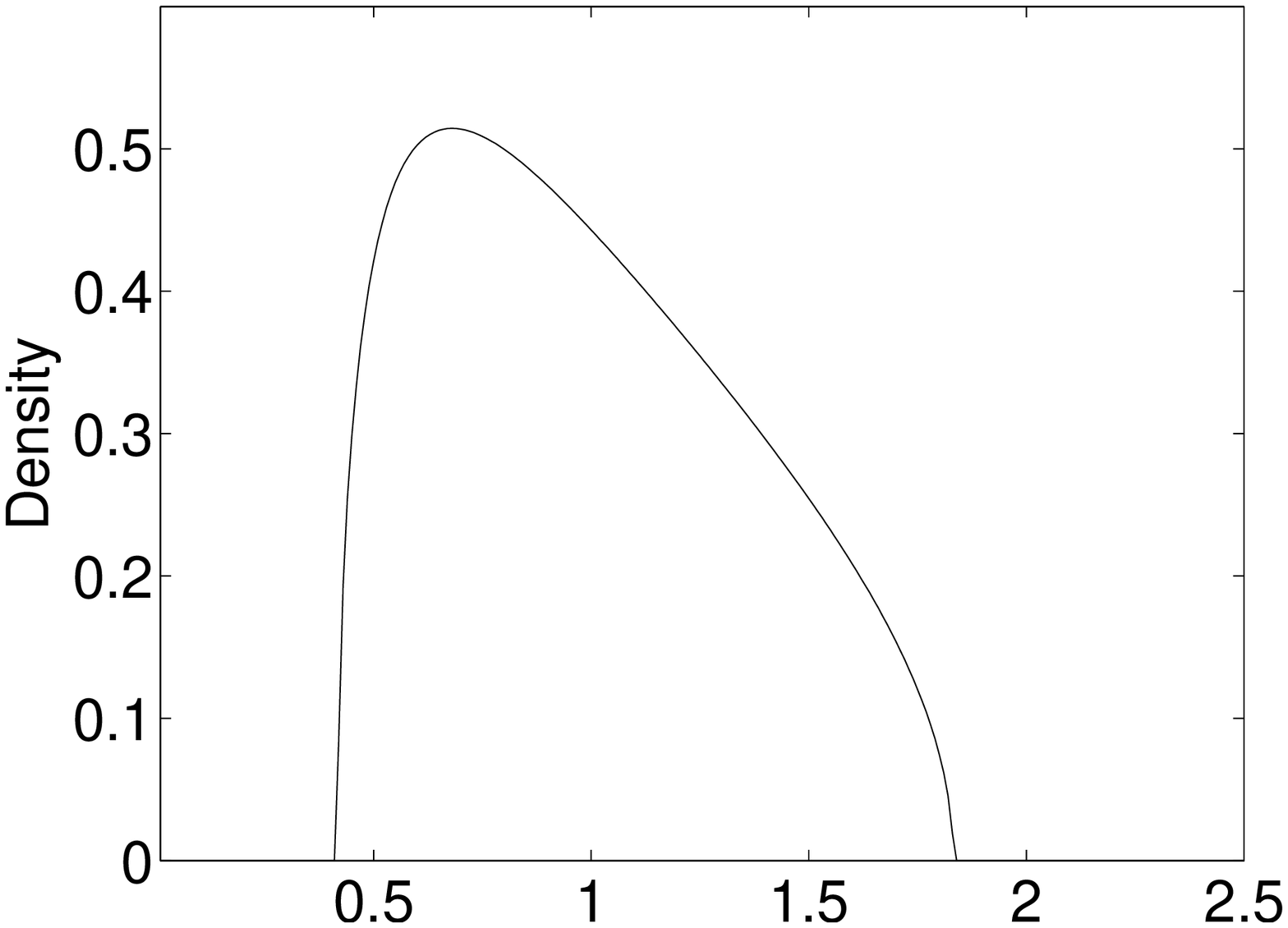,width=0.49\columnwidth}}

  \subfigure[$c=0.5$. $L=1024$ observations]{\epsfig{figure=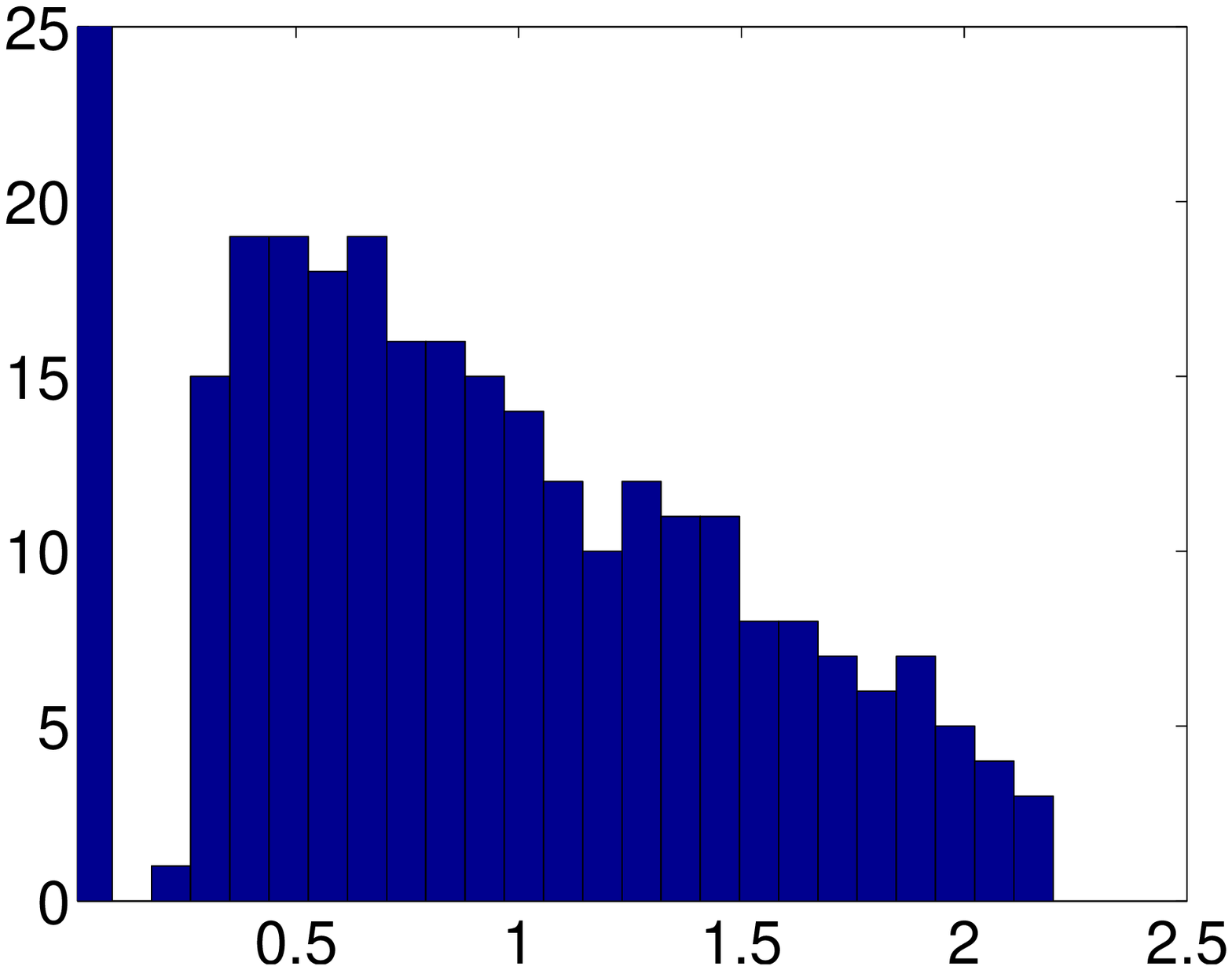,width=0.49\columnwidth}}
  \subfigure[$c=0.25$. $L=2048$ observations]{\epsfig{figure=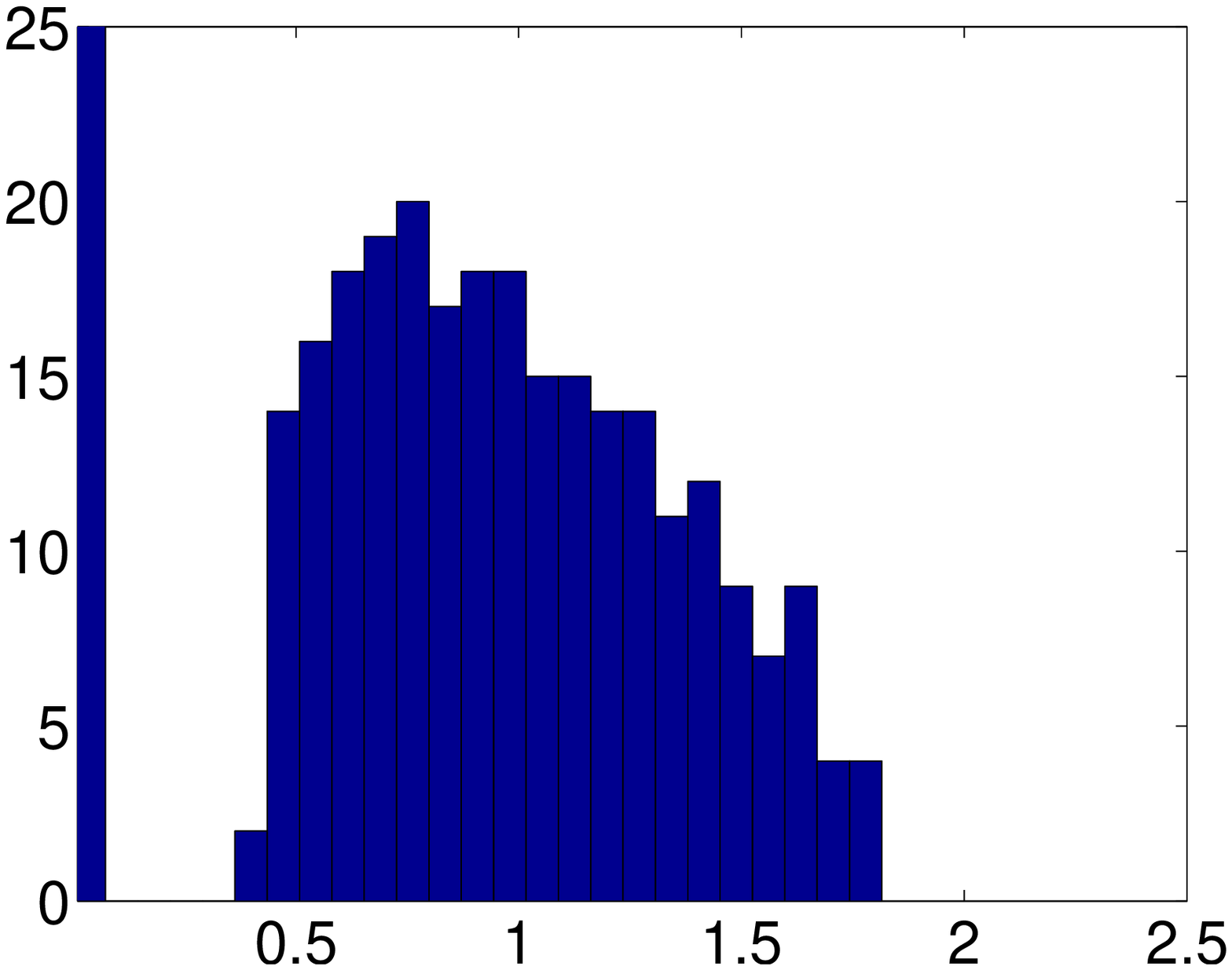,width=0.49\columnwidth}}
  \caption{Densities of $\left( \frac{1}{2}\delta_0 + \frac{1}{2}\delta_1 \right) \boxtimes\mu_c$ (upper row),
    and corresponding histogram of eigenvalues for the sample covariance matrices
    for different number of observations (lower row)}\label{fig:simexact}
\end{figure}

The importance of theorem~\ref{exactconv} is apparent: The mass of
$\mu\boxtimes\mu_c$ is seen to be centered on $\lambda(1+cp)$,
with support width of $4\lambda\sqrt{cp}$. If we let $c$ go to
zero, the center of mass approaches $\lambda$ and the support
width approaches zero. We note that the center of the support of
$\mu\boxtimes\mu_c$ is slighly perturbed to the right of
$\lambda$, while the density maximum occurs slightly to the left
of $\lambda$. It is easily checked that the support width and the
maximum density uniquely identifies a pair $(p,\lambda)$. This
means that, if we have an estimate  of the density of
$\mu\boxtimes\mu_c$ (for instance in the form of a realization of
a sample covariance matrix) for a measure $\mu$ of the form
(\ref{ourform}), the maximum density and the support width give us
a good candidate for the $(p,\lambda)$ defining $\mu$.
Figure~\ref{fig:simexact} shows densities of some realizations of
$\mu\boxtimes\mu_c$ for $p=\frac{1}{2}$ and $\lambda=1$, together
with corresponding realizations of covariances matrices. Values
$c=0.25$ and $c=0.5$ were used, with $L=1024$ and $L=2048$
observations respectively. Covariance matrices of size $512\times
512$ were used.

A similar result to theorem~\ref{exactconv} characterizing $\mu\boxslash\mu_c$ is proved in
appendix~\ref{sec:appendixb}:
\begin{theorem} \label{exactdeconv}
  The density of $\mu\boxslash\mu_c$ is $0$ outside the interval
  \begin{equation} \label{jcp}
      \begin{array}{lll}
        J_{\lambda , c , p} &=&[\lambda(1-2cp) - 2\lambda\sqrt{ cp(1-cp) } \\
                            &  & \lambda(1-2cp) + 2\lambda\sqrt{ cp(1-cp) }],
      \end{array}
  \end{equation}
  while the density on $J_{\lambda , c , p}$ is given by
  \begin{equation} \label{density2}
    f^{\mu\boxslash\mu_c}(x) = \frac{\sqrt{ L_1(x) L_2(x) }}{2cx^2},
  \end{equation}
  where
  \[
  \begin{array}{lll}
    L_1(x) &=& x - \lambda(1-2cp) + 2\lambda\sqrt{cp(1-cp)}\\
    L_2(x) &=& \lambda(1-2cp) + 2\lambda\sqrt{cp(1-cp)} - x.
  \end{array}
  \]
\end{theorem}

The support in this case is centered on $\lambda(1-2cp)$, which
is slightly to the left of $\lambda$, contrary to the case of
convolution. The support width is $4\lambda\sqrt{ cp(1-cp) }$.
Also in this case it is easily seen that the support narrows and
gets centered on $\lambda$, as $c$ goes to $0$. The densities in
(\ref{density}) and (\ref{density2}) are seen to resemble the
density of $\mu_c$ in (\ref{mpdensity}). One difference is that
$\mu_c$ is centered on $1$, while the densities in 
(\ref{density}) and (\ref{density2}) need not be.

The proofs of theorem~\ref{exactconv} and~\ref{exactdeconv} build on an analytical machinery for computing free convolution, where several transforms play a role.
Besides the Stieltjes transform $m_{\nu}(z)$, one has the $\eta$-transform, defined for real $z\geq 0$
\[
  \eta_{\nu}(z) = \int_{-\infty}^{\infty} \frac{1}{1+z\lambda}dF^{\nu}(\lambda).
\]
It is easy to verify that $m_{\nu}$ can be analytically continued to the negative part of the real line,
and that
\begin{equation} \label{sttoeta}
  \eta_{\nu}(z) = \frac{ m_{\nu}\left( -\frac{1}{z} \right) }{z}
\end{equation}
for $z\geq 0$.
The $\eta$-transform has some nice properties, it is for instance strictly monotone decreasing, so that it has an inverse.
In~\cite{eurecom:multfreeconv} it was shown that
\begin{equation} \label{origform}
  \eta_{\mu\boxtimes\mu_c}^{-1}(z) = \frac{ \eta_{\mu}^{-1}(z) }{1-c+cz}.
\end{equation}
The proof of this involved some more transforms, 
like the $S$-transform, and the actual value of these transforms for $\mu_c$.
The following forms will be more useful to us than (\ref{origform}), 
and can be deduced easily from it:
\begin{equation} \label{form1}
  \eta_{\mu} \left( z(1-c+c\eta_{\mu\boxtimes\mu_c}(z)) \right) = \eta_{\mu\boxtimes\mu_c}(z)
\end{equation}
and
\begin{equation} \label{form2}
  \eta_{\mu} \left( \frac{z}{1-c+c\eta_{\mu\boxslash\mu_c}(z)} \right) = \eta_{\mu\boxslash\mu_c}(z).
\end{equation}
$\eta_{\mu}(z)$ for $\mu$ as in (\ref{ourform}) is easily calculated:
\begin{equation} \label{etamueq}
  \eta_{\mu}(z) = 1 - p + \frac{p}{1+z\lambda}.
\end{equation}

$\mu\boxtimes\mu_c$ with unconnected support components centered
at the dirac locations of $\mu$ may very well happen for
discrete measures with more than two atoms also, but we do not
address this question here. The more general case, even when there
are two dirac's away from $0$, does not admit a closed-form
solution, since higher degree equations in general can not be
solved using algebraic methods. However, one can still solve those
equations numerically: For $\mu$ on the more general form
(\ref{ourforminit}), the $\eta$-transform is 
\[
  \eta_{\mu}(z) = \sum_{i=1}^n \frac{p_i}{1+z\lambda_i}.
\]
Putting this into (\ref{form1}), we see that we can solve
\[
  \sum_{i=1}^n \frac{p_i}{1+z\lambda_i \left( 1-c+c\eta_{\mu\boxtimes\mu_c}(z) \right)} = \eta_{\mu\boxtimes\mu_c}(z)
\]
to find $\eta_{\mu\boxtimes\mu_c}(z)$.
Collecting terms, we see that this is a higher order equation in $\eta_{\mu\boxtimes\mu_c}(z)$.
$m_{\mu}(z)$ and hence the density of $\mu$ can then be found from (\ref{sttoeta}).

\section{Applications to signal processing} \label{apps}
In this section, we provide several applications of free
deconvolution and show how the framework can be used in this
paper.

\subsection{Estimation of power and the number of users} \label{sim1}
In communication applications, one needs to determine the number
of users in a cell in a CDMA type network as well the power with
which they are received (linked to the path loss). Denoting by $n$
the spreading length, the received vector at the base station in
an uplink CDMA system is given by:
\begin{eqnarray}
{\bf y}_i={\bf W} {\bf P}^{\frac{1}{2}}{\bf s}_i+{\bf b}_i
\end{eqnarray}

where ${\bf y}_i$, ${\bf W}$, ${\bf P}$, ${\bf s}_i$  and ${\bf
b}_i$ are respectively the $n\times 1$ received vector, the
$n\times N$ spreading matrix with i.i.d zero mean, $\frac{1}{n}$
variance entries, the $N \times N$ diagonal power matrix,  the $N
\times 1$ i.i.d gaussian unit variance modulation signals and  the
$n\times 1$ additive white zero mean Gaussian noise.

Usual methods determine the power of the users by  finding  the
eigenvalues of  covariance matrix of ${\bf y}_i$ when the
signatures (matrix ${\bf W}$) and the noise variance are known.
\begin{eqnarray}
{\bf \Theta}={\mathbb{E}}\left({\bf y}_i {\bf y}_i^H\right)={\bf
W}{\bf P} {\bf W}^H+ \sigma^2 {\bf I}
\end{eqnarray}

However, in practice, one has only access to an estimate of the
covariance matrix and does not know the signatures of the users.
One can solely assume the noise variance known. In fact, usual
methods compute the sample covariance matrix (based on $L$
samples) given by:
\begin{eqnarray}
\hat{\bf \Theta}=\frac{1}{L}\sum_{i=1}^{L}{\bf y}_i{\bf y}_i^H
\end{eqnarray}

and determine the number of users  (and not the powers) in the
cell by the non zero-eigenvalues (or up to an ad-hoc threshold for
the noise variance) of:

\begin{eqnarray}
\hat{\bf \Theta}-\sigma^2 {\bf I}
\end{eqnarray}

This method, referred here as classical method, is quite
inadequate when $L$ is in the same range as $n$. Moreover, it does
not provide a method for the estimation of the power of the users.

The free deconvolution framework introduced in this paper is well
suited for this case and enables to determine the power of the
users without knowing their specific code structure. Indeed, the
sample covariance matrix is related to the true covariance matrix
${\bf \Theta}={\mathbb{E}}\left({\bf y}_i {\bf y}_i^H\right)$ by:

\begin{eqnarray} \label{use1}
\hat{\bf \Theta}={\bf \Theta}^{\frac{1}{2}} {\bf X}{\bf X}^H {\bf
{\bf \Theta}^{\frac{1}{2}}}
\end{eqnarray}

with
\begin{eqnarray} \label{use2}
{\bf \Theta}={\bf W}{\bf P} {\bf W}^H + \sigma^2 {\bf I}
\end{eqnarray}
and ${\bf X}$ is a $n \times L$ i.i.d Gaussian zero mean matrix.

Combining (\ref{use1}), (\ref{use2}),
with the fact that ${\bf W}^H {\bf W}$, $\frac{1}{L} {\bf X}{\bf X}^H$ are Wishart matrices
with distributions approaching $\mu_{\frac{N}{n}}$, $\mu_{\frac{n}{L}}$ respectively, and using that
\[
  \mu_{ {\bf W}{\bf P} {\bf W}^H } =
    \frac{N}{n}\mu_{ {\bf W}^H{\bf WP}} + \left( 1 - \frac{N}{n}
    \right) \delta_0,
\]
we get due to asymptotic freeness the equation
\begin{equation} \label{sim1solve}
  \left(
    \left(
      \frac{N}{n} ( \mu_{\frac{N}{n}} \boxtimes \mu_{{\bf P}} ) + \left( 1 - \frac{N}{n} \right) \delta_0
    \right)
    \boxplus \mu_{\sigma^2 I}
  \right)
  \boxtimes \mu_{\frac{n}{L}}
  =
  \mu_{\bf \hat{R}}
\end{equation}
If one knows the noise variance, one can use this equation in
simulations in two ways:
\begin{enumerate}
  \item Through additive and multiplicative free deconvolution, use (\ref{sim1solve})
  where
    the power distribution of the users (and de facto the number of users) is expressed in terms of the sample covariance matrices.
  \item Determine the numbers of users $N$ through a best-match
    procedure: Try all values of $N$ with $1\leq N\leq n$,
    and choose the $N$ which gives a best match between the left and right hand side in (\ref{sim1solve}).
\end{enumerate}
To solve (\ref{sim1solve}), method A was used to compute the moments. In order to
solve (\ref{step1}), we also need to compute additive free
deconvolution with a scalar. This was addressed in
section~\ref{strategy1}, but can also be computed in a simpler
way,  since
\[
  \phi( (a-\sigma^2I)^j ) = \sum_{k=0}^j (-1)^k \left( \stackrel{j}{k} \right) \sigma^{2k} \phi(a^{j-k}).
\]
In (\ref{sim1solve}) we also scale a measure with $\frac{N}{n}$, and
add an atom at $0$. Both of these cases are easily implemented.

In the following simulations, a spreading length of $n=256$ and
noise variance $\sigma^2=0.1$ have been used.

\subsubsection{Estimation of power}
We use a $36 \times 36$ ($N=36$) diagonal matrix as our power
matrix ${\bf P}$, and use three sets of values, at 0.5, 1 and 1.5
with equal probability, so that
\begin{equation} \label{sim1acdf}
  \mu_{\bf P} = \frac{1}{3} \delta_{0.5} + \frac{1}{3} \delta_{1} + \frac{1}{3} \delta_{1.5}.
\end{equation}
There are no existing methods for estimating such a $\mu_{\bf P}$
from the sample covariance matrices: to our knowledge, existing
methods estimate the power with non-zero eigenvalues of the sample
covariance matrix up to $\sigma^2$. In our case, the powers are
all above $\sigma^2$.

\begin{figure}
  \subfigure[$L=256$]{\epsfig{figure=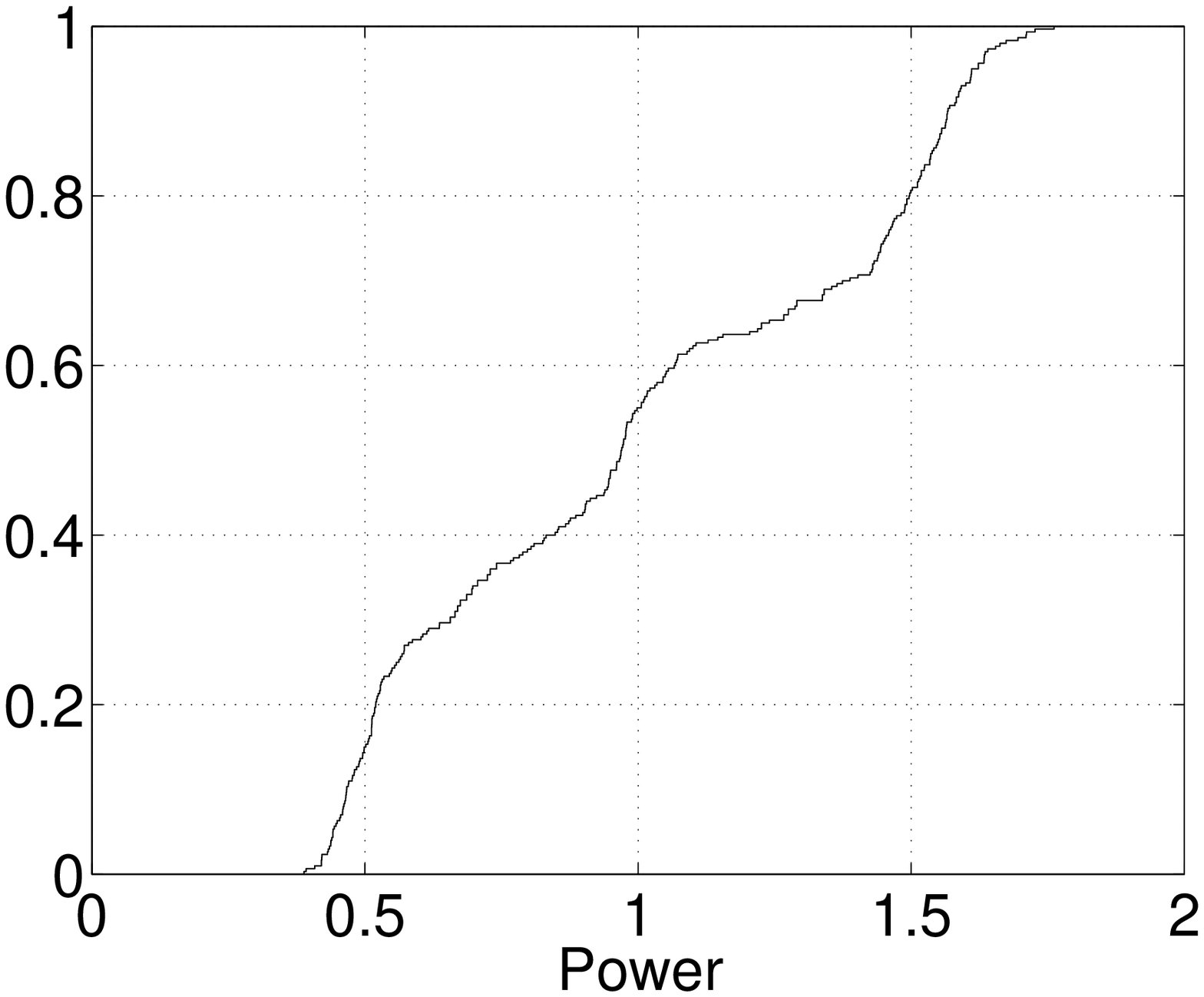,width=0.49\columnwidth}}
  \subfigure[$L=512$]{\epsfig{figure=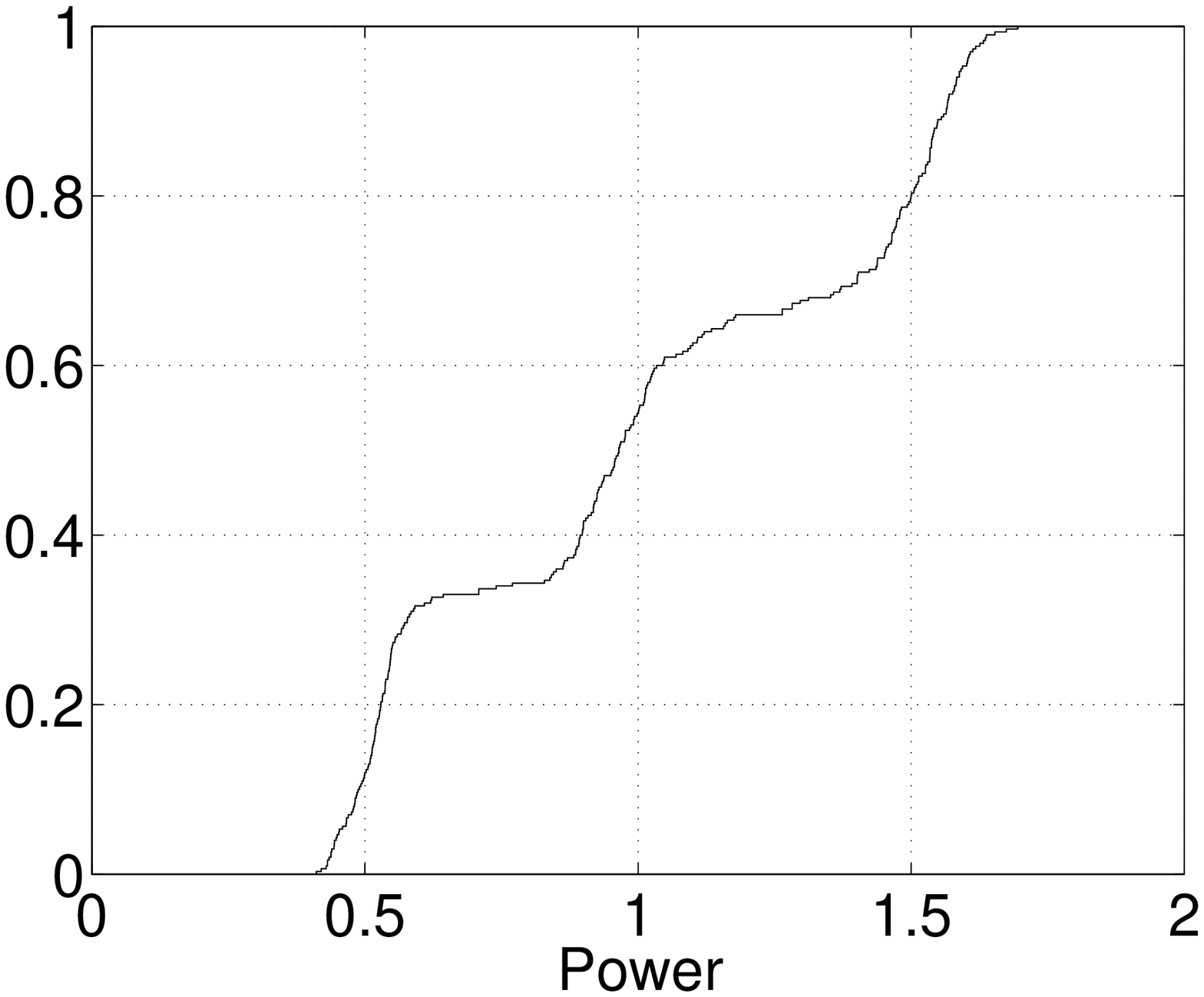,width=0.49\columnwidth}}

  \subfigure[$L=1024$]{\epsfig{figure=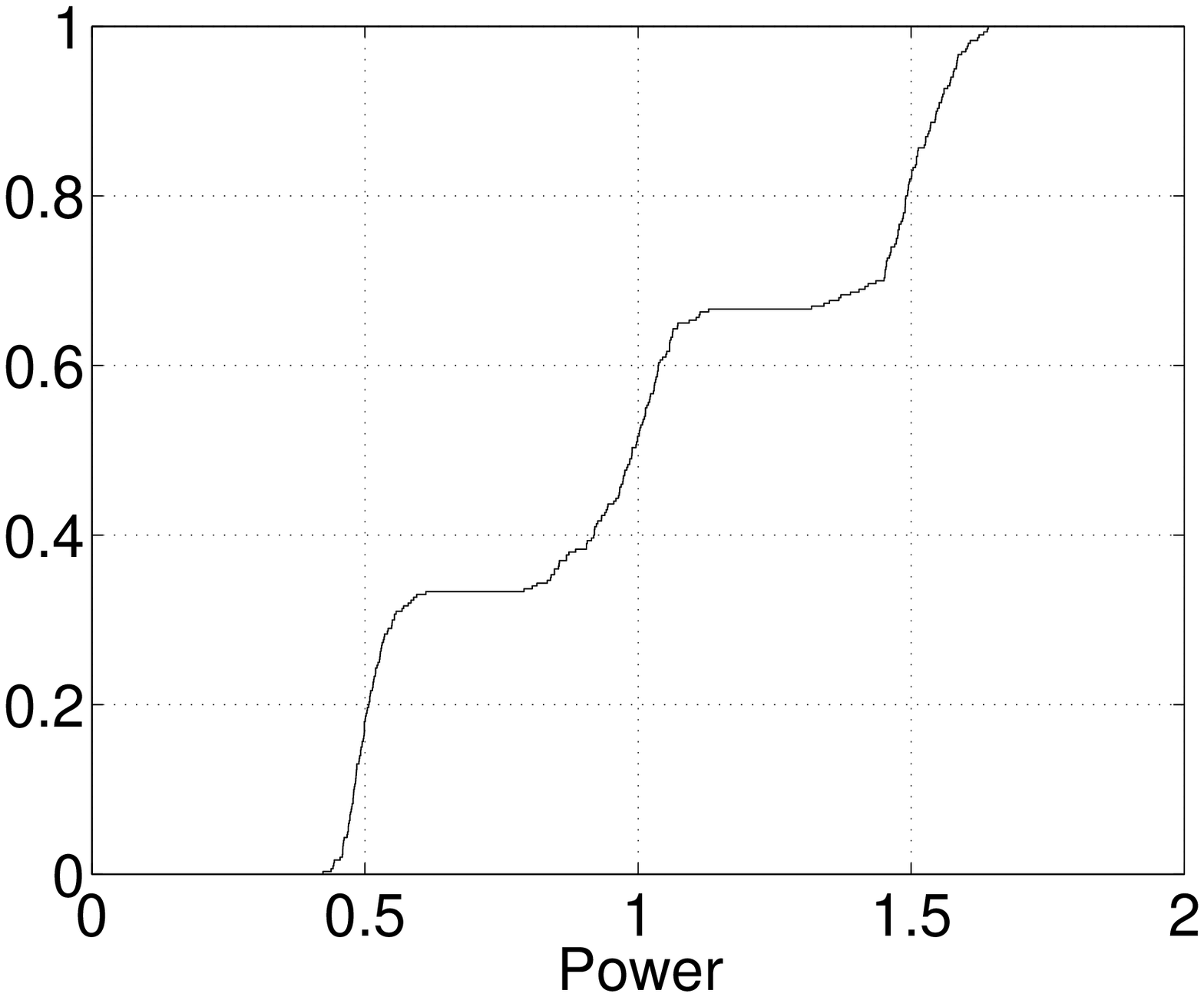,width=0.49\columnwidth}}
  \subfigure[$L=2048$]{\epsfig{figure=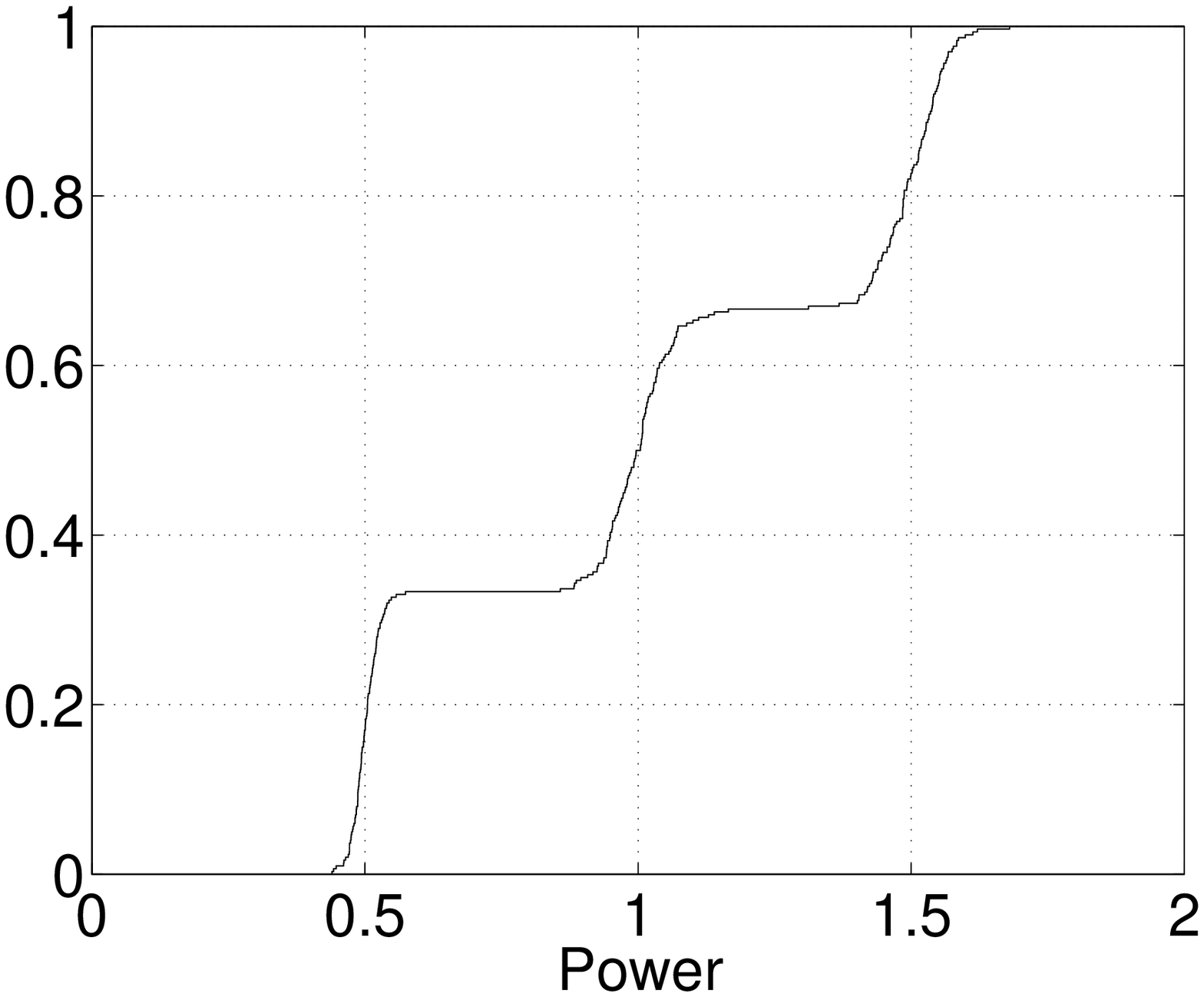,width=0.49\columnwidth}}
  \caption{CDF of powers estimated from multiplicative free deconvolution from sample covariance matrices with different number of observations.}\label{fig:sim1acdf}
\end{figure}

\begin{figure}
  \begin{center}
    \epsfig{figure=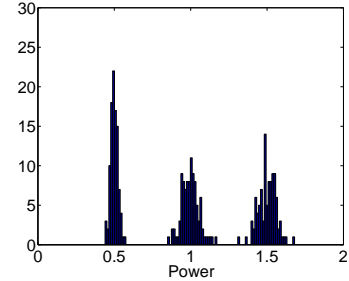,width=0.5\columnwidth}
  \end{center}
  \caption{Distribution of the powers estimated from multiplicative free deconvolution from sample covariance matrices with $L=2048$.}\label{fig:sim1ahist}
\end{figure}

In figure~\ref{fig:sim1acdf}, the CDF of $\mu_{\bf P}$ was
estimated by solving (\ref{sim1solve}), using method A with three
moments. The resulting moments from method A were used to compute
estimates of the eigenvalues through the Newton-Girard formula,
and the CDF was computed by averaging these eigenvalues for 100
runs for each number of observations. An alternative strategy
would be to use higher moments, and less runs for each
observation. As one can see, when L increases, we get a CDF closer
to that of (\ref{sim1acdf}). The best result is obtained for
$L=2048$. The corresponding histogram of the eigenvalues in this
case is shown in figure~\ref{fig:sim1ahist}.

\subsubsection{Estimation of the number of users}
We use a $36 \times 36$ ($N=36$)  diagonal matrix as our power
matrix ${\bf P}$ with  $\mu_{\bf P} =  \delta_{1}$.  In this case,
a common method that try to find just the rank exists. This method
determines the number of eigenvalues greater than $\sigma^2$. Some
threshold is used in this process. We will set the threshold at
$1.5\sigma^2$, so that only eigenvalues larger that $1.5\sigma^2$
are counted. There are no general known rules for where the
threshold should be set, so some guessing is inherent in this
method. Also, choosing a wrong threshold can lead to a need for a
very high number of observations for the method to be precise.

We will compare this classical method with a free convolution method for estimating the rank, following the procedure sketched in step 2).
The method is tested with varying number of observations, from $L=1$ to $L=4000$,
and the $N$ which gives the best match with the moments of the SCM in (\ref{sim1solve}) is chosen.
Only the four first moments are considered.
\begin{figure}
  \begin{center}
    \epsfig{figure=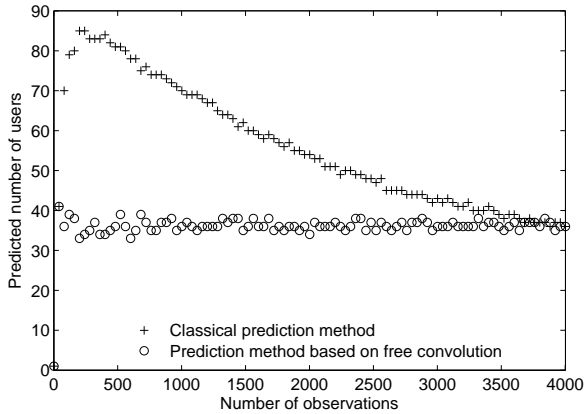,width=0.99\columnwidth}
  \end{center}
  \caption{Estimation of the number of users with a classical method, and free convolution $L=1024$ observations have been used.}\label{fig:sim1b}
\end{figure}
In figure~\ref{fig:sim1b}, it is seen that when $L$ increases, we
get a prediction of $N$ which is closer to the actual value $36$. 
The classical method starts to predict values close to the
right one only for a number of observations close to $4000$.
The method using free probability predicts values close to the
right one for a less greater number of realizations.

\subsection{Estimation of Channel correlation} \label{sim2}
In channel modelling, the modeler would like to infer on the
correlation between the different degrees of the channel. These
typical cases are represented by a  received signal (assuming that
a unit training sequence has been sent) which is given by
\begin{eqnarray}
{\bf y}_i={\bf w}_i+{\bf b}_i
\end{eqnarray}

where ${\bf y}_i$, ${\bf w}_i$ and ${\bf b}_i$ are respectively
the $n\times 1$ received vector, the $n\times 1$ zero Gaussian
impulse response and $n\times 1$ additive white zero mean Gaussian
noise with variance $\sigma$. The cases of interest can be:
\begin{itemize}
\item Ultra-wide band
applications~\cite{paper:telatartse,paper:porrattsenacu,paper:turinjanaghassemzadehricetarokh,paper:netomenounidebbahfleury}
where one measures in the
  frequency domain the wide-band nature of the frequency signature
  ${\bf w}_i$
\item Multiple antenna applications
\cite{paper:telatar99,ieeeJSAC.Gesbert03} with one
  transmit and $n$ receiving antennas where ${\bf w}_i$ is the spatial
  channel signature at time instant $i$.
\end{itemize}

Usual methods compute the sample covariance matrix given by:
\begin{eqnarray*}
\hat{\bf R}=\frac{1}{L}\sum_{i=1}^{L}{\bf y}_i{\bf y}_i^H
\end{eqnarray*}

The sample covariance matrix is related to the true covariance
matrix of ${\bf w}_i$ by:

\begin{eqnarray}
\hat{\bf R}={\bf \Theta}^{\frac{1}{2}} {\bf X}{\bf X}^H {\bf {\bf
\Theta}^{\frac{1}{2}}}
\end{eqnarray}

with
\begin{eqnarray}
{\bf \Theta}={\bf R} +\sigma^2 {\bf I}
\end{eqnarray}

and ${\bf X}$ is an $N \times n$ i.i.d Gaussian zero mean matrix.

Hence, if one knows the noise variance (measured without any
signal sent), one can determine the eigenvalue distribution of the
true covariance matrix following:

\begin{equation} \label{sim2solve}
  \mu_{\bf R} = (\mu_{\bf \hat{R}} \boxslash \mu_{\frac{n}{L}}) \boxminus \mu_{\sigma^2 I}.
\end{equation}
According to theorem~\ref{teo2}, computing $\mu_{\bf \hat{R}}
\boxslash \mu_{\frac{n}{L}}$ is the same as computing the
$G_2$-estimator for covariance matrices. Additive free
deconvolution with $\mu_{\sigma^2 I}$ is the same as performing a
shift of the spectrum to the left.

We use a rank $K$ covariance matrix of the form ${\bf
R}=diag[1,1,..,1,0,..,0]$, and variance $\sigma^2 = 0.1$, so that $\sigma\sim 0.3162$. For simulation purposes,  $L$  vectors
${\bf w}_i$ with covariance {\bf R}  have been generated with
$n=256$ and $K=128$. We would like to observe the p.d.f.
\begin{equation} \label{sim2cdf}
  \frac{1}{2} \delta_0 + \frac{1}{2} \delta_1
\end{equation}
in our simulations.

\begin{figure}
  \subfigure[$L=128$]{\epsfig{figure=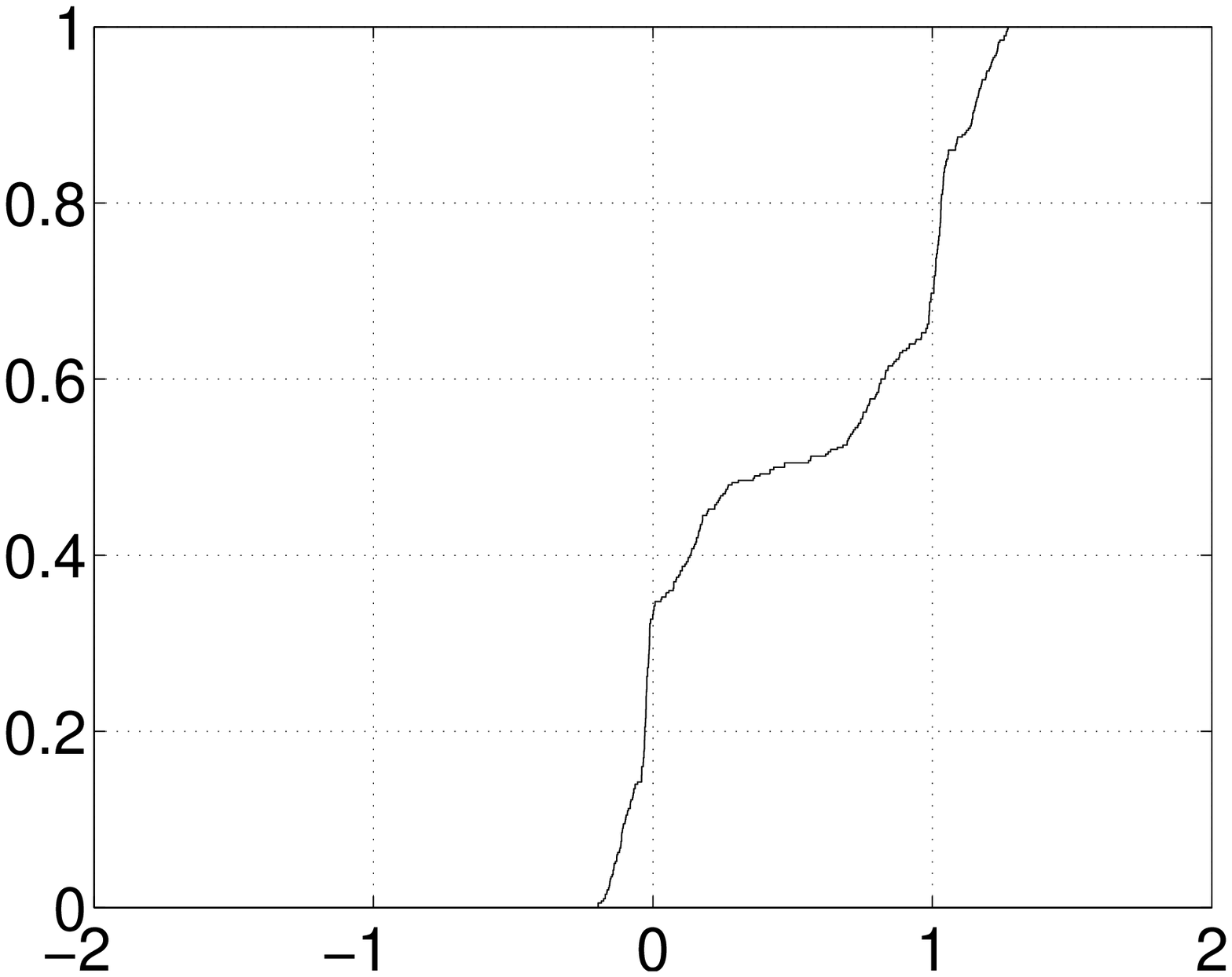,width=0.49\columnwidth}}
  \subfigure[$L=512$]{\epsfig{figure=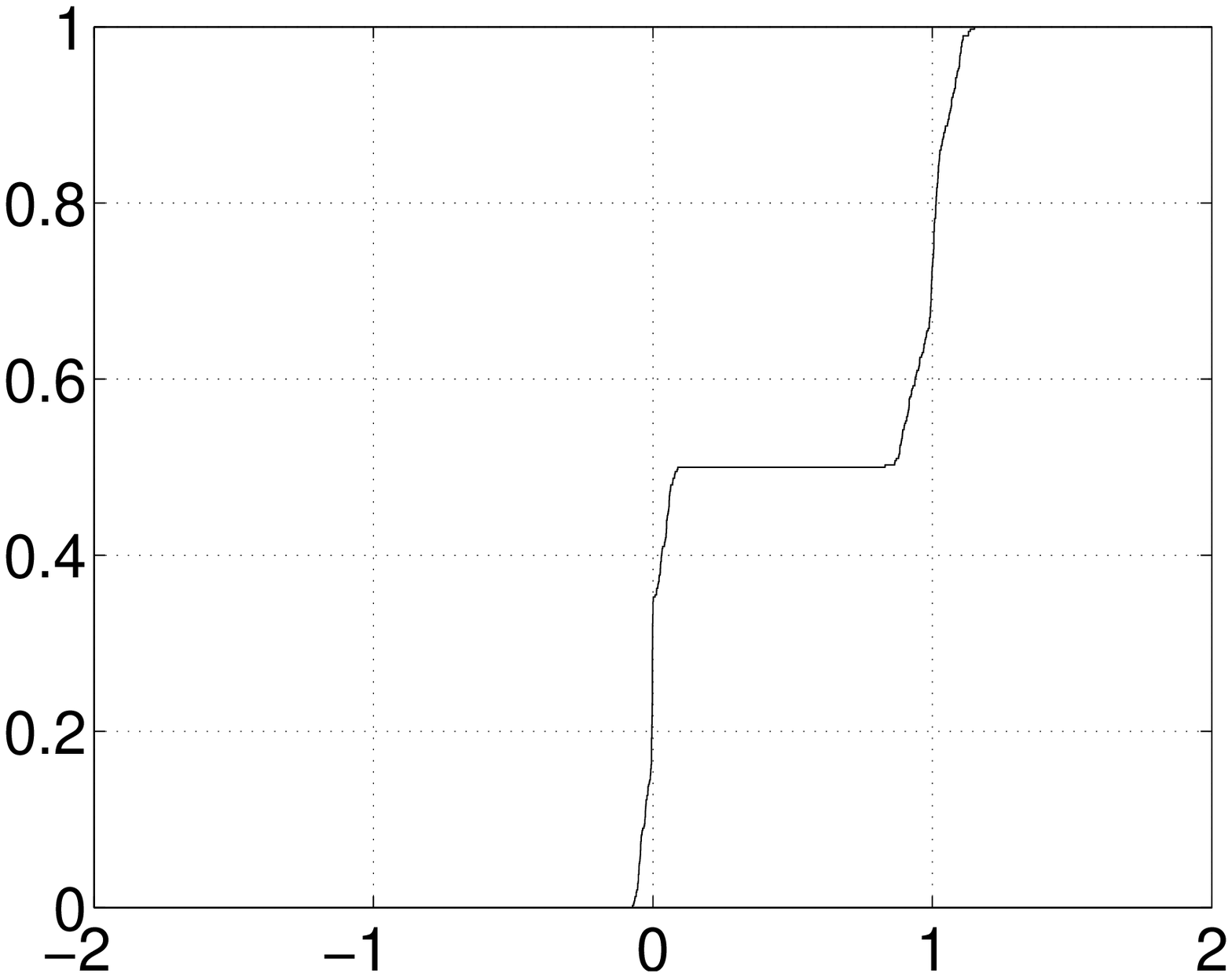,width=0.49\columnwidth}}
  \caption{CDF of eigenvalues estimated from multiplicative free deconvolution from sample covariance matrices with different number of observations.}\label{fig:sim2acdf}
\end{figure}

 In figure~\ref{fig:sim2acdf}, (\ref{sim2solve}) has been solved,
using $L=128$ and $L=512$ observations, respectively. 
The same strategy as in section~\ref{sim1} was used, i.e. the CDF was produced by averaging eigenvalues from 100 runs. 
4 moments were computed. 
Both cases suggest a p.d.f. close to that of (\ref{sim2cdf}).
It is seen that the number of observations need not be higher than the dimensions of the systems in order for free deconvolution to work.

The second case corresponds to $c=0.5$, so that when there is no
noise ($\sigma = 0$), the sample covariance is approximately 
$\left( \frac{1}{2}\delta_0 + \frac{1}{2}\delta_1 \right) \boxtimes\mu_{\frac{1}{2}}$, 
which is shown in figure~\ref{fig:simexact} a). If it is known
that the covariance has the density $(1-p)\delta_0 +
p\delta_{\lambda}$, theorem~\ref{exactconv} can be used, so that
we only need to read the maximum density and the location
parameter in order to find $p$ and $\lambda$.

%

\begin{figure}
  \begin{center}
    \epsfig{figure=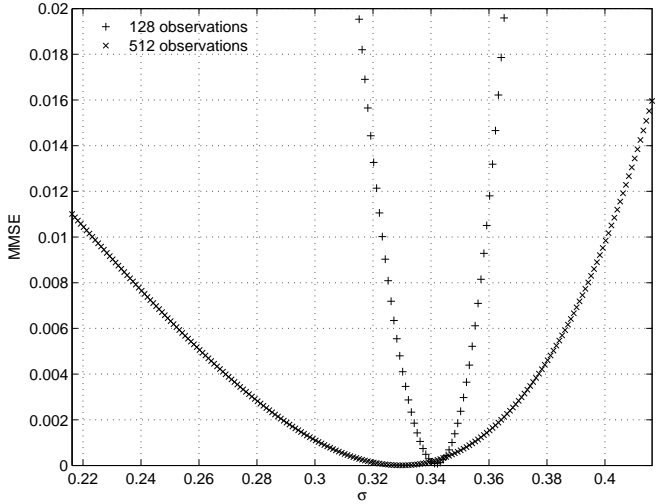,width=0.99\columnwidth}
  \end{center}
  \caption{Estimation of the noise variance. $L=128$ and $L=512$ observations have been used.}\label{fig:sim2b}
\end{figure}

It may also be that the true covariance matrix is known,
and that we would like to estimate the noise variance through a limited number of observations.
In figure~\ref{fig:sim2b}, $L=128$ and $L=512$ observations have been taken.
In accordance with (\ref{sim2solve}), we compute $(\mu_{\bf R} \boxplus \mu_{\eta^2 I}) \boxtimes \mu_{\frac{n}{L}}$ for
a set of noise variance candidates $\eta^2$, and an MSE of the four first moments of this with the moments of the observed sample covariance matrix is computed.
Values of $\eta$ in $(\sigma - 0.1,\sigma + 0.1)\sim (0.2162,0.4162)$ have been tested, with a spacing of 0.001.
It is seen that the MMSE occurs close to the value $\sigma = \sqrt{0.1} = 0.3162$, even if the number of observations is smaller than the rank.
The MMSE occurs closer to $\sigma$ for $L=512$ than for $L=128$, so the estimate of sigma improves slightly with $L$. 
It is also seen that the MSE curve for $L=512$ lies lower than the MSE curve for $L=128$.
An explanation for this lies in the free convolution with $\mu_{\frac{n}{L}}$:
As $L\rightarrow\infty$, this has the effect of concentrating all energy at $1$.

\section{Further work}
In this work, we have only touched upon a fraction of  the
potential of free deconvolution in the field of signal processing.
The framework is well adapted for any problem where one needs to
infer on one of the mixing matrices. Moreover, tools were
developed to practically deconvolve the measures numerically and
where shown to be simple to implement. Interestingly, although the
results are valid in the asymptotic case, the work presented in
this paper shows that it is well suited for sizes of interest for
signal processing applications. The examples draw upon some basic
wireless communications problems but can be extended to other
cases. In particular, classical blind methods
\cite{donoho,comon,belouchrani} which assume an infinite number of
observations or noisyless problems can be revisited in light of
the results of this paper. The work can also be extended in
several directions and should bring new insights on the potential
of free deconvolution.

\subsection{Other applications to signal processing}
There are many other examples that could be considered in this
paper. Due to limitations, we have detailed only two. For example,
another case of interest can be the estimation of channel
capacity.  In usual measurement methods, one validates models
\cite{ieeeJSAC.Kermoal02,paper.ozcelik03,vtc.ozcelik05,paper.debbah06}
by determining how the model fits with actual capacity
measurements. In this setting, one has to be extremely cautious
about the measurement noise.

Indeed, the MIMO measured channel is given by:
\begin{eqnarray}
  \hat{{\bf H}}_i = \frac{1}{\sqrt{n}} \left( {\bf  H} + \sigma {\bf X}_i \right)
\end{eqnarray}

where $\hat{\bf H}_i$,  ${\bf H}$ and  ${\bf X}_i$ are
respectively the $n \times n$ measured MIMO matrix ($n$  is the
number of receiving and transmitting antennas), the $n \times n$
MIMO channel and the $n \times n$ noise matrix with i.i.d zero
mean unit  variance  Gaussian entries. We suppose that the channel
stays constant (block fading assumption) during $L$ blocks. In
this case, the observed model becomes:

\begin{eqnarray}
  \hat{{\bf H}}_{1...L} = \frac{1}{\sqrt{n}} \left( {\bf  H}_{1...L} + \frac{\sigma}{\sqrt{L}} {\bf X}_{1...L} \right)
\end{eqnarray}

with

\begin{eqnarray}
  \hat{{\bf H}}_{1...L} =\frac{1}{\sqrt{L}} \left[\hat{{\bf H}}_1, \hat{{\bf
  H}}_2,...,\hat{{\bf H}}_L\right]
\end{eqnarray}

\begin{eqnarray}
 {\bf  H}_{1...L} = \frac{1}{\sqrt{L}} \left[{\bf  H},{\bf  H},...,{\bf  H}\right]
\end{eqnarray}

\begin{eqnarray}
 {\bf X}_{1...L} =   \left[{\bf X}_{1},{\bf X}_{2},...,{\bf X}_L\right]
\end{eqnarray}

The capacity of a channel with channel matrix ${\bf H}$ and signal
to noise ratio $\rho=\frac{1}{\sigma^2}$ is given by

\begin{eqnarray}
C &=& \frac{1}{n}\log \det\left( {\bf I}+\frac{1}{n \sigma^2} {\bf H}{\bf H}^{H} \right).\\
&=& \frac{1}{n} \sum_{l=1}^{n} \log(1+ \frac{1}{\sigma^2}
\lambda_l)
\end{eqnarray}

where $\lambda_l$ are the eigenvalues of $\frac{1}{n} {\bf H}{\bf
H}^{H}$. The problem consists therefore of estimating the
eigenvalues of $\frac{1}{n} {\bf H}{\bf H}^{H}$ based on few
observations   of $\hat{\bf H}_i$. For a single observation, This
can be done through the approximation
\begin{equation} \label{decthis}
  \mu_{ \hat{\bf H}_1\hat{\bf H}_1^{H} } \boxslash \mu_1 =
  \left( \mu_{\frac{1}{n} {\bf H}{\bf H}^{H}} \boxslash \mu_1 \right) \boxplus \mu_{\sigma}.
\end{equation}
If we have many observations, we have that:

\begin{equation} \label{decthis2}
  \mu_{ \hat{\bf H}_{1...L}\hat{\bf H}_{1...L}^{H} } \boxslash \mu_1 =
  \left( \mu_{\frac{1}{n} {\bf H}{\bf H}^{H}} \boxslash \mu_1 \right) \boxplus \mu_{\frac{\sigma}{\sqrt{L}}}.
\end{equation}

\subsection{Other types of sample matrices}
One topic of interest is the use of free deconvolution with other
types of matrices than the sample covariance matrix. In fact, based
on a given set of observations, one can construct higher sample
moment matrices than the sample covariance matrix (third product
matrix for example). These matrices contain useful information
that could be used in the problem. The main difficult issue here
is to prove freeness of the convolved measures. The free
deconvolution framework could also be applied to tensor problems
\cite{kruskal} and this has not been considered yet to our
knowledge.

\subsection{Colored Noise}
In this work, the noise considered was supposed to be temporally
and spatially white with standard Gaussian entries. This yields
the Mar\u{c}henko pastur law as the operand measure. However, the
analysis can be extended, with the assumption that freeness is
proved, to other type of noises: the case for example of an
additive noise with a given correlation. In this case, the operand
measure is not the Mar\u{c}henko pastur law but depends on the
limiting distribution of the sample noise covariance matrix.
Although the mathematical formulation turns out to be identical,
in terms of implementation, the problem is more complicated as one
has to use more involved power series than the $Zeta$-series for
example.

\subsection{Parametrized distribution}
In the previous example (signal impaired with noise), the
Mar\u{c}henko Pastur law $\mu_c$ was one of the operand measures,
while the other  was either estimated or considered to be a
discrete measure, i.e. with density
\begin{equation} 
  f^{\mu}(x) = \sum_{i=1}^n p_i\delta_{\lambda_i}(x).
\end{equation}

 It turns out that one can find also the parameterized
distribution (best fit by adjusting the parameter) that
deconvolves up  to certain minimum mean square error. For example,
one could approximate the measure of interest with two diracs
(instead of the set of $n$ diracs) and find the best set of diracs
that minimizes the mean square error. One can also approximate the
measure with  the Mar\u{c}henko pastur law for which the parameter
$c$ needs to be optimized. In both cases, the interesting point
is that the expressions can be derived explicitly.

\section{Conclusion}
In this paper, we have   shown that free probability provides a
neat framework for estimation problems when the number of
observations is of the same order as the dimensions of the
problem. In particular, we have introduced a free deconvolution
framework (both additive and multiplicative) which is very
appealing from a mathematical point of view and provides an
intuitive understanding of some G-estimators provided by Girko
\cite{chapter:girkotenyears}. Moreover, implementation aspects
were discussed and proved to be adapted, through simulations, to
classical signal processing applications without the need of
infinite dimensions.

\appendices
\section{Algorithm for computing free convolution} \label{sec:appendix0}
The files {\tt momcum.m} and {\tt cummom.m} in~\cite{eurecom:freeimpl} are implementations
of (\ref{computableform}) in MATLAB. The first calculates cumulants from moments,
the second moments from cumulants. Both programs are rather short, and
both take a series of moments $(\mu_1,...,\mu_n)$ as input.
The algorithm for computing cumulants from moments goes the following way:
\begin{enumerate}
  \item Form the vector $m = (1,\mu_1,...,\mu_n)$ of length $n+1$,
    and compute and store recursively the $n$ vectors
    \[
      M_1 = m \mbox{, } M_2 = m\star m \mbox{,..., } M_n = \star_n m,
    \]
    where $\star_n$ stands for $n$-fold (classical) convolution with itself.
    The later steps in the algorithm use only the $n+1$ first elements of the vectors $M_1$, $M_2$,...,$M_n$.
    Consequently, the full $M_k$ vectors are not needed for all $k$:
    We can truncate $M_k$ to the first $n+1$ elements after each convolution,
    so that the implementation can be made quite efficient.
  \item Calculate the cumulants recursively.
    If the first $n-1$ cumulants, i.e. the first $\alpha_i$ in (\ref{computableform}), have been found by solving the $n-1$ first equations in
    (\ref{computableform}), $\alpha_n$ can then be found through the
    $n$th equation in (\ref{computableform}), by using the vectors computed in step 1).
    More precisely ,the connection between the vectors in 1) and the value we use in (\ref{computableform}) is
    \[
      coef_{n-k}\left( \left( 1+\mu_1 z + \mu_2z^2 +...\right)^k \right) = M_k(n-k),
    \]
    where $n-k$ denotes the index in the vector (starting from $0$).
    Finding the $k$'th cumulant $\alpha_k$ by solving the $k$th equation in (\ref{computableform}) is the same as
    \[
      \alpha_k = \frac{ M_1(n+1) - \sum_{1\leq r\leq k-1} \alpha_r M_r(k-r) }{ M_k(0) }.
    \]
\end{enumerate}
The program for computing moments from cumulants is slightly more complex, since we can't start out by computing the vectors
$M_1$,...,$M_n$ separately at the beginning, since the moments are used to form them (these are not known yet).
Instead, elements in $M_1$,...,$M_n$ are added each time a new moment has been computed.

\section{The proof of theorem~\ref{exactconv}} \label{sec:appendixa}
Set $\eta(z) = \eta_{\mu\boxtimes\mu_c}$. From (\ref{form1}) we see that we must solve the equation
\[
  \eta_{\mu} \left( z(1-c+c\eta(z)) \right) = \eta(z).
\]
Substituting (\ref{etamueq}), multiplying and collecting terms, we get that $\eta(z)$ must be a zero for the equation
\[
  cz\lambda \eta(z)^2 + \left( 1 + z\lambda (1-2c+cp)) \right) \eta(z) -(1-p)(1-c)z\lambda - 1.
\]
The analytical continuation of $m(z) = m_{\mu\boxtimes\mu_c}(z)$ to the negative part of the real line satisfies $\eta(z) = \frac{ m\left( -\frac{1}{z} \right) }{z}$.
Subsituting this and also substituting $u=-\frac{1}{z}$, we get that
\begin{equation} \label{gconveq}
  -c\lambda z m(z)^2  + (\lambda(1-2c+cp)-z)m(z) + \frac{1}{z}\lambda (1-p)(1-c) - 1
\end{equation}
equals $0$ for $z$ which are real and negative.
It is clear that any analytical continuation of $m(z)$ to the upper half of the complex plane also must satisfy (\ref{gconveq}).
We use the formula for the solution of the second degree equation and get that $m(z)$ equals
\begin{equation} \label{neweq}
  \begin{array}{ll}
          &  \frac{ -\lambda(1-2c+cp) + z \pm\sqrt{ \begin{array}{l} (\lambda(1-2c+cp)-z)^2\\ +4c\lambda^2 (1-p)(1-c) - 4c\lambda z\end{array} }}{-2c\lambda z}\\
          &= \frac{  (\lambda(1-2c+cp)+z \mp
             \sqrt{ \begin{array}{l} z^2 - 2\lambda(1+cp)z \\ + 4c\lambda^2(1-c)(1-p) \\ + \lambda^2(1-2c+cp)^2 \end{array} }}{2c\lambda z}.
  \end{array}
\end{equation}
The zeroes of the discriminant here are
\[
  \begin{array}{lll}
      & & \frac{ 2(\lambda(1+cp) \pm \sqrt{ \lambda^2 4(1+cp)^2 - 16c\lambda^2(1-c)(1-p) - 4\lambda^2(1-2c+cp)^2} }{2}\\
      &=& \lambda(1+cp) \\
      & & \pm \frac{1}{2} \lambda\sqrt{ \begin{array}{l} 4(1+cp)^2 - 16c(1-c)(1-p) \\ -4(1+cp)^2 - 16c^2 + 16c(1+cp) \end{array} }\\
      &=& \lambda(1+cp) \pm \frac{1}{2} \lambda\sqrt{ 16c (1+cp -c -(1-c)(1-p))}\\
      &=& \lambda(1+cp) \pm 2\lambda\sqrt{cp},
  \end{array}
\]
This means that we can rewrite $m(z)$ to
\[
  \frac{\lambda(1-2c+cp)+z \mp\sqrt{ \begin{array}{l} (z - \lambda(1+cp) + 2\lambda\sqrt{cp})\\(z - \lambda(1+cp) - 2\lambda\sqrt{cp})\end{array} } }{2c\lambda z}.
\]
Thus, for $z$ real, $m(z)$ is complex if and only if $z$ lies in the interval $I_{\lambda , c , p}$ of (\ref{icp}).
Outside $I_{\lambda , c , p}$, the density of $\mu\boxtimes\mu_c$ is zero.
Taking the imaginary part and using the Stieltjes inversion formula, we get that the density in $I_{\lambda , c , p}$ is given by the formula (\ref{density}).

Setting the derivative of (\ref{density}) w.r.t. $z$ equal to $z$ gives us a first degree equation which yields a unique maximum at $z=\lambda\frac{(1-cp)^2}{1+cp}$.
After some more calculations, we get that the density at this extremal point is $\frac{\sqrt{cp}}{c\pi\lambda(1-cp)}$.
This finishes the proof.
\sluttmerke

\section{The proof of theorem~\ref{exactdeconv}} \label{sec:appendixb}
Set $\eta(z) = \eta_{\mu\boxslash\mu_c}$, we see from (\ref{form2}) that we must solve the equation
\[
  \eta_{\mu} \left( \frac{z}{1-c+c\eta(z)} \right) = \eta(z).
\]
Substituting (\ref{etamueq}), multiplying and collecting terms, we get that $\eta(z)$ must be a zero for the equation
\[
  c\eta(z)^2 + (1-2c+z\lambda)\eta(z) -(1-c)-z\lambda (1-p).
\]
Using the formula for the solution of the second degree equation, one can see that the positive square root must be chosen whenever $c<\frac{1}{2}$,
since $\eta$ assumes positive positive values whenever $z\geq 0$.
Substituting $\eta(z) = \frac{ m\left( -\frac{1}{z} \right) }{z}$ and also $u=-\frac{1}{z}$ as in the case for convolution, we get that
\[
  cz^2m(z)^2 + (\lambda - z(1-2c))m(z) + \frac{1}{z}\lambda (1-p) - (1-c)
\]
equals $0$ for $z$ which are real and negative. We get that $m(z)$ equals
\begin{equation}
  \begin{array}{ll}
    &  \frac{-\lambda + z(1-2c) \pm\sqrt{ \begin{array}{l} (\lambda - z(1-2c))^2 \\ - 4cz^2\left( \frac{1}{z}\lambda (1-p) - (1-c) \right) \end{array} }}{2cz^2}\\
    &= \frac{-\lambda + z(1-2c) \pm\sqrt{ z^2 - 2\lambda(1-2cp)z+\lambda^2 }}{2cz^2}.
  \end{array}
\end{equation}
The zeroes of the discriminant are
\[
  \begin{array}{ll}
     & \frac{ 2\lambda(1-2cp) \pm\sqrt{ 4\lambda^2(1-2cp)^2 -4\lambda^2} }{2}\\
    =& \lambda(1-2cp) \pm\frac{1}{2}\sqrt{ 4\lambda^2(4c^2p^2-4cp)^2 }\\
    =& \lambda(1-2cp) \pm 2\lambda\sqrt{ cp(1-cp)}.
  \end{array}
\]
Following the same reasoning as for convolution, we see that the density is $0$ outside the interval $J_{\lambda , c , p}$ of (\ref{jcp}),
and that the density in $J_{\lambda , c , p}$ is given by (\ref{density2}).
This finishes the proof.
\sluttmerke

\bibliography{pti}

\end{document}